\documentclass[aps,article,author-year,notitlepage]{revtex4-1}
\usepackage{subeqn}
\usepackage{graphicx}
\usepackage{float}
\usepackage[T1]{fontenc}
\usepackage[latin1]{inputenc}
\usepackage{amssymb}
\include{epsf}
\epsfverbosetrue

\newcommand{\bea}{\begin{eqnarray}}
\newcommand{\eea}{\end{eqnarray}}
\newcommand{\ts}{\vspace*{.2cm}}	

\begin{document}
\large

\title{A new method to calculate the $n$-Particle Irreducible Effective Action}

\author{M.E. Carrington}
\email{carrington@brandonu.ca}
\affiliation{Department of Physics, Brandon University, Brandon, Manitoba, R7A 6A9 Canada\\ and \\  Winnipeg Institute for Theoretical Physics, Winnipeg, Manitoba, Canada }
\author{Yun Guo}
\email{guoyun@brandonu.ca}
\affiliation{Department of Physics, Brandon University, Brandon, Manitoba, R7A 6A9 Canada\\ and \\  Winnipeg Institute for Theoretical Physics, Winnipeg, Manitoba, Canada }

\begin{abstract}

In this paper, we present a new method to calculate the $n$-Loop $n$-particle irreducible effective action.
The key is an organizational trick that involves the introduction of a set of fictitious bare vertices that are set to zero at the end of the calculation. Using these fictitious vertices, we prove that the Schwinger-Dyson equations are the same as the equations of motion obtained from the $n$-particle irreducible  effective action, up to the level at which they respect the symmetries of the original theory. This result allows us to obtain the effective action directly from the Schwinger-Dyson equations, which are comparatively easy to calculate.
As a check of our method, we reproduce the known results for the $n$-Loop $n$-particle irreducible effective action with $n=4$ and $n=5$. We also use the technique to calculate the 6-Loop 6-particle irreducible effective action.

\end{abstract}

\pacs{03.65.Ud, 03.67.Mn}
\maketitle

\section{Introduction}

An $n$-Loop $n$-particle irreducible ($n$PI)  effective theory is defined in terms of $n$ functional arguments which correspond to a set of $n$-point functions that are determined self-consistently through a variational procedure. The idea was introduced in Refs. \cite{deDom1,deDom2} and first discussed in the context of relativistic field theories in  Ref.  \cite{norton}. The variational procedure resums certain classes of diagrams, and represents a reorganization of perturbation theory. $n$PI approximation schemes are especially interesting because they can be used to study far-from-equilibrium systems \cite{cox,aartsNonEq0,smitNonEq,bergesNonEq,aartsNonEq1,aartsNonEq2}, which is of interest in the context of heavy ion collisions and cosmology. The potential importance of $n$PI theories is demonstrated by the fact that they can be used to formulate the calculation of transport coefficients \cite{edQED,edQCD,EK4}. To date however, numerical calculations have only been done for 2PI theories where it has been shown that the convergence of perturbative approximations is improved  (see \cite{bergesReview,bergesSEWM2004,bergesConvg} and references therein).
In addition, there are unresolved issues for gauge theories \cite{calzettaReview,julienReview}. The renormalizability of a theory is related to the existence of symmetry constraints on the $n$-point functions. For $n$PI effective theories, symmetries and renormalizabilty are connected to the fact that proper $n$-point functions can be defined in more than one way. All definitions are completely equivalent for the exact theory, but they are not the same at finite approximation order.
These issues are well understood for scalar theories and QED at the 2PI level.
For scalar theories, one can define a 2-point function that satisfies Goldstone's theorem in the broken phase \cite{baier2PI,vanHees3}. For QED, one can define $n$-point functions that obey Ward identities \cite{edQED-2pi,julienSym,calzettaSym}. These symmetry constraints allow one to construct a renormalized theory that preserves the symmetries of the original theory \cite{vanHees1,vanHees2,vanHees3,reinosaRenorm1,reinosaRenorm2}.
For non-Abelian theories, the situation is more involved. It has been shown that at any order in the approximation scheme, the gauge dependence of the effective action always appears at higher approximation order \cite{smit,HZ}. However, the gauge symmetries of the $n$-point functions are more complicated than for Abelian theories, and renormalizability remains an open question.

In this paper, we introduce a new method to calculate the $n$PI effective action.
While it is true in principle that the effective action can always be obtained from a series of Legendre transforms, this method is extremely complicated for $n>3$, and probably prohibitively tedious beyond $n=5$. The  3-Loop 4PI effective action was calculated in Refs. \cite{deDom1,deDom2,MEC,berges}, the 4-Loop 4PI effective action in Ref. \cite{EK4}, and the 5-Loop 5PI effective action in Ref. \cite{paper1}.
The key to our method is the introduction of a set of fictitious bare vertices: to obtain the $n$-Loop $n$PI effective action we include in the Lagrangian the vertices $V_j^{oo}$ for $j=3,4,5,6,\cdots\,,n$. The inclusion of the nonrenormalizable interactions ($j\ge 5$) is an organizational trick, and these vertices will be set to zero at the end of the calculation.
Using these fictitious vertices, we can show that the equations of motion (eom's) and Schwinger-Dyson (sd) equations are equivalent to the order at which the truncated theory respects the symmetries of the original theory.
This result allows us to construct the $n$-Loop $n$PI effective action directly from the sd equations.

This paper is organized as follows. In Sec. \ref{section:notation}, we define our notation. In Sec. \ref{structure}, we discuss the basic structure of the $n$PI effective action. In Sec. \ref{sdeom}, we prove that the eom's and sd equations are equivalent to the truncation order. Our new method to calculate the effective action is explained in detail in Sec. \ref{newApproachSection}.
In Secs. \ref{example44} and \ref{example55}, we show how to reproduce, with comparatively little effort, the known results for the $n$-Loop $n$PI effective action with $n=4$ and $n=5$. This provides a check of the procedure. In Sec. \ref{example66},  we use the technique to calculate the 6-Loop 6PI effective action which is, realistically speaking, impossibly tedious to obtain using Legendre transforms.

We make one further comment. Our method is based on the fact that using fictitious vertices in intermediate steps of the calculation, the sd equations can be rewritten so that they have the same structure as the eom's. It is important to realize that this result is important {\it only} because it allow us to obtain the effective action without taking a series of Legendre transforms. It is {\it not} true that the nonperturbative solutions of a truncated set of sd equations are the same as the solutions of the eom's obtained from the $n$PI effective action.

\section{Notation}
\label{section:notation}

Throughout this paper we use $L$ to indicate the loop order in the skeleton expansion. We also use ``$n$-Loop'' to mean terms in the skeleton expansion with $L\le n$ loops, and ``$n$-loop'' to mean terms in the skeleton expansion with $L =  n$ loops. We consider only scalar theories. The generalization of the method to other theories is straightforward.

In most equations in this paper, we suppress the arguments that denote the space-time dependence of functions. As an example of this notation, the quadratic term in the action is written [see Eq. (\ref{scl})]:
\bea
\label{notex}
\frac{1}{2}\int d^4 x\, d^4y\,\varphi(x)\big[i (D^{oo})^{-1}(x-y)\big]\varphi(y)~~ \rightarrow ~~ \frac{i}{2}(D^{oo})^{-1}\varphi^2\,.
\eea
We define several different kinds of vertex functions and use the letter $V$ for all of them, with a single subscript denoting the number of legs:
\bea
\label{vertDefs}
&&V_j^{oo}~~\text{bare vertex - equation (\ref{scl})}\, , \\
&&V_j^{0}~~\text{effective bare vertex - equation (\ref{oo20})}\, ,\nonumber \\
&&V_j^c~~\text{connected vertex - equation (\ref{Wders})}\, ,\nonumber \\
&&V_j~~\text{proper vertex - equation (\ref{properDefn})}\, ,\nonumber \\
&&\tilde V_j~~\text{tilde vertex - equation (\ref{tildeDef})}\, .\nonumber
\eea
Unless stated otherwise, the indices $\{j,k,l,\cdots\}$, which indicate the number of legs on a bare, effective bare, connected, proper, or tilde vertex, run from 3 to $n$. In diagrams, bare vertices and proper vertices are denoted by open circles and solid dots, respectively\footnote{Figures in this paper are drawn using Jaxodraw \cite{jaxo}.}. Many of the equations we will write in this paper are easier to understand as diagrams. In some cases, we will give only the diagrammatic form of an equation.

To illustrate a limitation of our notation, we write the equation that relates the 4-point connected vertex to proper vertices without suppressing space-time arguments. We use a single index to denote a space-time variable, and the summation convention to mean integration. The standard result is
\bea
\label{proper}
V^c_{ijkl}&& =D_{i t_1} D_{j t_2}
   D_{k t_3} D_{l t_4}
   V_{t_1 t_2 t_3 t_4}+D_{it_1}D_{j t_2} D_{k t_3}
   D_{l t_4}D_{t_5 t_6}
   V_{t_1 t_6 t_3}
   V_{t_2 t_5 t_4}\\
&&+D_{i t_1} D_{j t_2}
   D_{k t_3} D_{l t_4} D_{t_6 t_5}
   V_{t_1 t_2 t_6}
   V_{t_3 t_5 t_4}+D_{i t_1} D_{j t_2}
   D_{k t_3} D_{l t_4} D_{t_6 t_5}
   V_{t_1 t_5 t_4}
   V_{t_6 t_2 t_3}\,.\nonumber
\eea
Using our notation in which indices are suppressed, the distinction between the $s$, $t$, and $u$ channels is lost and the second, third, and fourth terms on the right side become $(3)D^5 V_3^2$. We indicate that all three channels are included in one term by writing the factor (3) in brackets. In all calculations, contributions to a given vertex that correspond to different permutations of external legs must be treated correctly. The abbreviated notation only allows us to present results in a simpler form.

We introduce some terminology for different types of graphs that could appear in the effective action.

\begin{figure}[H]
\begin{center}
\includegraphics[width=14cm]{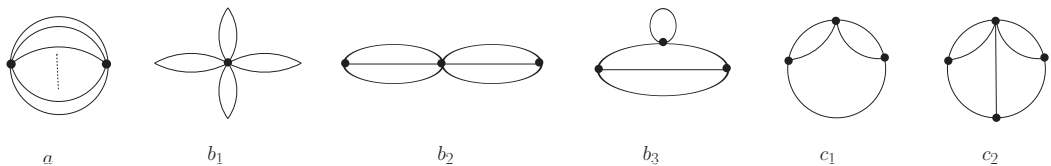}
\end{center}
\caption{\label{superExLABEL}Some of graphs that could appear in the effective action. In graph ($b_1$), the 8-point vertex is the tadpole vertex. In graph ($b_2$), the 6-point vertex is the tadpole vertex. In graph ($c_1$), the 4-point vertex is the flower vertex. In graph ($c_2$), the 5-point vertex is the flower vertex.}
\end{figure}

\begin{description}


\item[Basketballs]
Graphs with two $V_{j}$ vertices which are connected by $j$ propagators. A generic example is shown in part (a) of Fig. \ref{superExLABEL}.
\item[Tadpoles]
Graphs that would produce disconnected contributions to the equation of motion of one of the vertices in the graph (which we call the ``tadpole vertex''). Some examples are shown in part (b) of Fig. \ref{superExLABEL}.  Tadpole graphs with only 1 vertex [for example, part ($b_1$) in Fig. \ref{superExLABEL}] are type (1), and all other tadpole graphs [for example,  part ($b_2$) in Fig. \ref{superExLABEL}] are type (2).

\item[Flowers]
Graphs that would produce nonproper (1PR) contributions to the equation of motion of one of the vertices in the graph (which we call the ``flower vertex''). Some examples are shown in part (c) of Fig. \ref{superExLABEL}.

\end{description}

\par

The effective action is calculated using a trick which involves introducing a set of fictitious bare vertices as an organizational tool. At the end of the calculation, the bare vertices are set to zero for $j\ge 5$. The classical action is
\bea
\label{scl}
S_{cl}[\varphi]=\frac{1}{2}\varphi \big[i\,(D^{oo})^{-1}\big]\varphi-\sum_{j=3}^n\frac{i}{\;j!} V_j^{oo}\varphi^j\,.
\eea
It will be useful to define an effective bare propagator and effective $j$-point vertex as
\bea
\label{free}
&&(D^0(\phi))^{-1}=-i \frac{\delta^2 S_{cl}[\phi]}{\delta \phi^2}\,,~~~V^0_j(\phi)=i\frac{\delta^j S_{cl}[\phi]}{\delta \phi^j}\,.
\eea
From now on, we suppress the argument and write $D^0(\phi) \to D^0$ and $V^0_j(\phi) \to V^0_j$.
The general relation between bare vertices $V_j^{oo}$ and effective bare vertices $V_j^{0}$ is
\bea
\label{oo20}
V_l^0 = \sum^n_{j=l} \frac{1}{(j-l)!}V_j^{oo}\phi^{j-l}\,.
\eea

\section{Structure of the Effective Action}
\label{structure}

The $n$PI effective action is defined as the $n$th Legendre transformation of the connected generating functional which is constructed by coupling the field to $n$ source terms:
\bea
\label{genericGamma}
&& Z[R_j]=\int d\varphi  \;{\rm Exp}[i\,(S_{cl}[\varphi]+\sum_{j=1}^n\frac{1}{j!} R_j\varphi^j)]\,,\\[1mm]
&&W[R_j]=-i \,{\rm Ln} Z[R_j]\,,\nonumber\\[1mm]
&&\Gamma[\phi,D,V^0_j,V_k] = W -\sum_{j=1}^n R_j\frac{\delta W}{\delta R_j} \,.\nonumber
\eea
The last line in (\ref{genericGamma}) gives the effective action as an implicit function of effective bare and proper vertices.
We define connected green functions:
\bea
\label{Wders}
V_j^c = \langle\varphi^j\rangle_c = -(-i)^{j+1}\frac{ \delta^j W}{\delta R_1^j}\,.
\eea
The equations that relate the connected and proper vertices are obtained from their definitions using the chain rule\footnote{Equations (\ref{Wders}) and (\ref{properDefn}) are also valid for $j=1,\;2$. Equation (\ref{properDefn}) gives $V_2=D^{-1}$ and thus $\Gamma[V_1,V_2,V_3\dots V_n]$ really means $\Gamma[\phi,D^{-1},V_3,V_4,\dots]$ and not $\Gamma[\phi,D,V_3,V_4,\dots]$. We ignore this point to avoid introducing unnecessary notation.}
\bea
\label{properDefn}
V_j= i \frac{\delta^j}{\delta \phi^j} \Gamma_{\rm 1PI} = i \frac{\delta^j}{\delta \phi^j}\left(W[R_1]-R_1 \phi\right)\,.
\eea
We organize the calculation of the effective action using the
method of subsequent Legendre transforms \cite{berges,MEC}. This method involves starting from an expression for the  2PI effective action and exploiting the fact that the source terms $R_j$ for $j\ge 3$ can be combined with the corresponding bare vertices by defining a set of modified interaction vertices which we call tilde vertices:
\bea
\label{tildeVert}
\label{tildeDef}
\tilde{V}_j:= V^{oo}_j +i R_{j}\,.
\eea
Using these tilde vertices, we can rewrite the effective action in (\ref{genericGamma}) as
\bea
\label{genericGamma2}
\Gamma[\phi,D,V^{0}_j,V_k] =: \tilde\Gamma_{\rm 2PI} - \sum_{j=3}R_j\frac{\delta W}{\delta R_j}\,.
\eea
We will refer to $\tilde\Gamma_{\rm 2PI}$ as the tilded 2PI effective action. It is constructed from the complete set of $n$-Loop 2PI diagrams for a theory with bare vertices $V^{oo}_j$ ($3\le j\le n)$ by replacing all bare vertices with tilde vertices\footnote{The only role of the fictitious vertices in a calculation using subsequent Legendre transforms is to introduce 2PI diagrams containing vertices $V^{oo}_j$ for $5\le j\le n$, which are then replaced by tilde vertices.}.

The 2PI effective action has the form
\bea
\label{gammaGen2PI}
&&\Gamma_{\rm 2PI}[\phi,D,V_j^{oo}] =S_{cl}[\phi]+\frac{i}{2} {\rm Tr} \,{\rm Ln}D^{-1}  +
\frac{i}{2} {\rm Tr}\,\big[ \left(D^0\right)^{-1} D\big]-i\Phi[\phi,D,V_j^{oo}]~~+~~{\rm const} \,,
\eea
where $\Phi[\phi,D,V_j^{oo}]$ contains all contributions to the effective action with two or more loops.
It is convenient to divide the 0-loop and 1-loop contributions to $\tilde\Gamma_{\rm 2PI}$ into pieces that do and do not contain tilde vertices:
\bea
\label{listDef}
&& \tilde\Gamma_{\rm 2PI}:=\Gamma_{\rm 2PI}[\phi,D,\tilde V_j] = \Gamma^{oo}_0+\tilde\Gamma_0+\Gamma^{oo}_1 +\tilde\Gamma_1 -i\tilde\Phi \,,\\
&& \Gamma^{oo}_0=\frac{i}{2} (D^{oo})^{-1} \phi^2 \text{~contains 0-loop graphs with no tilde vertices}\,,\nonumber \\
&& \tilde\Gamma_0 = -\sum_{j=3}^n\frac{i}{j!}\tilde V_j\phi^j \text{~contains 0-loop graphs with tilde vertices}\,,\nonumber\\
&& \Gamma^{oo}_1 = \frac{i}{2} {\rm Tr}\big((D^{oo})^{-1}\,D\big)+\frac{i}{2}{\rm Tr}\,{\rm Ln}D^{-1}
\text{~contains 1-loop graphs with no tilde vertices}\,,\nonumber\\
&& \tilde\Gamma_1=\frac{i}{2}{\rm Tr}\big[\big((\tilde D^0)^{-1} - (D^{oo})^{-1}\big)D\big] \text{~contains 1-loop graphs with tilde vertices}\,,\nonumber\\
&& \tilde\Phi = \Phi[\phi,D,\tilde V_j]\,.\nonumber
\eea

Using Eqs. (\ref{genericGamma}), (\ref{Wders}), (\ref{tildeDef}), and (\ref{genericGamma2}), we have
\bea
\label{side1}
\langle \phi^j\rangle && = j!\frac{\delta W}{\delta R_j} = j!\frac{\delta\tilde\Gamma_{\rm 2PI}}{\delta R_j} = i j! \frac{\delta\tilde\Gamma_{\rm 2PI}}{\delta\tilde V_j}\, \\
&& =: V_j^c+ \chi_j\,.\nonumber
\eea
The term $\chi_j$ contains all disconnected contributions to the expectation value, and is a function of connected vertices. A general expression for these terms is given in Appendix \ref{yunFormula}.
Substituting (\ref{tildeDef}) and (\ref{side1}) into (\ref{genericGamma2}), we have
\bea
\label{dormant2}
\Gamma[\phi,D,V^{0}_j,V_k]  =  \tilde\Gamma_{\rm 2PI}+ \sum_{j=3}^n \frac{i}{j!}(V^c_j+\chi_j)(\tilde V_j-V_j^{oo})\,.
\eea
Using (\ref{listDef}), Eq. (\ref{dormant2}) becomes
\bea
\label{dormant2point2}
\Gamma[\phi,D,V^{0}_j,V_k]
&&=  \Gamma^{oo}_0+\tilde\Gamma_0+ \Gamma^{oo}_1+\tilde\Gamma_1 \\
&& -i\big(\tilde\Phi_{\rm basketball} + \tilde\Phi^{(1)}_{\rm tadpole} + \tilde\Phi^{(2)}_{\rm tadpole}+ \tilde\Phi_{\rm flower} + \tilde\Phi_{\rm rest}\big) \nonumber\\
&& +\sum_{j=3}^n\frac{i}{j!}\bigg(V^c_j + \chi^{(0)}_j + \chi^{(1)}_j + \chi^{(2)}_j + \chi^{(3)}_j\bigg)(\tilde V_j-V^{oo}_j)\,, \nonumber
\eea
where we have separated contributions from the different types of graphs in $\tilde\Phi$ as discussed in Sec. \ref{section:notation}.
The terms $\chi_j^{(i)}$ for $i$=0,1,2 refer to specific pieces of $\chi_j$ and $\chi_j^{(3)}$ includes all other contributions.
We give some examples that will be useful in the discussion below (see Appendix \ref{yunFormula}):
\bea
\label{chiExamples}
&&j=3: ~~  \chi^{(0)}_3 = \phi^3\,,~~~\chi^{(1)}_3 = (3)D\phi\,,~~~\chi^{(2)}_3 = \chi_3^{(3)} = 0 \,,\\
&&j=4: ~~  \chi^{(0)}_4 = \phi^4\,,~~~\chi^{(1)}_4 = (6)D\phi^2\,,~~~\chi^{(2)}_4 = (3)D^2\,,~~~\chi_4^{(3)} = (4)V_3^c \phi\,,\nonumber \\
&&j=5:~~  \chi^{(0)}_5 = \phi^5\,,~~~\chi^{(1)}_5 = (10)D\phi^3\,,~~~\chi^{(2)}_5 = (15)D^2\phi\,,~~~\chi_5^{(3)} = (5)V_4^c \phi+(10)V_3^c\phi^2+(10)V_3^c D\,.\nonumber
\eea

\ts

We define a functional that contains all terms in $\Gamma[\phi,D,V^{0}_j,V_k]$ that have bare vertices:
\bea
\label{gamma0Def}
\Gamma^0[\phi,D,V^0_j,V_k]&&:=\Gamma^{oo}_0+\Gamma^{oo}_1-\sum_{j=3}^n\frac{i}{j!}(V^c_j+\chi_j)V_j^{oo}\,.
\eea
The right side of (\ref{gamma0Def}) does not contain tilde vertices, and it is straightforward to convert connected vertices to proper ones, and bare vertices to effective bare vertices. The result has the form
\bea
\label{gamma0Def2}
\Gamma^0[\phi,D,V^0_j,V_k]=:S_{cl}[\phi]+ \frac{i}{2} {\rm Tr} \,{\rm Ln}D^{-1}  +
\frac{i}{2} {\rm Tr}\,\big[ \left(D^0\right)^{-1} D\big] -i \Phi^0[V_j^0,V_k]\,,
\eea
where $\Phi^0[V_j^0,V_k]$ contains all diagrams with more than one loop.
The procedure is discussed in detail in Appendix \ref{bareSection}.

\ts

Using (\ref{chiDef}) and the definitions in Eq. (\ref{listDef}), it is straightforward to show
\bea
\label{can1}
\tilde\Gamma_0+\sum_{j=3}^n\frac{i}{j!}\chi_j^{(0)}\tilde V_j = 0\,,\\
\tilde\Gamma_1+ \sum_{j=3}^n\frac{i}{j!} \chi_j^{(1)}\tilde V_j = 0\,,\nonumber\\
-i\tilde\Phi^{(1)}_{\rm tadpole} +\sum_{j=3}^n \frac{i}{j!}\chi^{(2)}_j\tilde V_j = 0\,.\nonumber
\eea

Substituting (\ref{gamma0Def}) and (\ref{can1}) into (\ref{dormant2point2}), we have
\bea
\label{dormant2point3}
&& \Gamma[\phi,D,V^{0}_j,V_k] \\
&&=  \Gamma^0[\phi,D,V^0_j,V_k] -i \big(\tilde\Phi_{\rm basketball}+\tilde\Phi^{(2)}_{\rm tadpole}+ \tilde\Phi_{\rm flower}+ \tilde\Phi_{\rm rest}\big) +\sum_{j=3}^n\frac{i}{j!}(V_j^c  + \chi^{(3)}_j) \tilde V_j \,. \nonumber
\eea

Equation (\ref{dormant2point3}) is a formal result for the effective action as an implicit function of proper vertices. The right side is a function of tilde vertices and connected vertices.
Comparing (\ref{dormant2point3}) with (\ref{dormant2}), we have
\bea
\tilde\Gamma_{\rm 2PI} = -i \big(\tilde\Phi_{\rm basketball}+\tilde\Phi^{(2)}_{\rm tadpole}+ \tilde\Phi_{\rm flower}+ \tilde\Phi_{\rm rest}\big)
 + \sum_{j=3}^n\frac{i}{j!}\big(\chi_j^{(3)}-\chi_j\big)\,\tilde V_j + \cdots
 \eea
where the dots represent terms that do not depend on $\tilde V_j$. Substituting into (\ref{side1}), we obtain
\bea
\label{unwrap2}
V^c_j= j! \frac{\delta}{\delta \tilde V_j}(\tilde\Phi_{\rm basketball}+\tilde\Phi^{(2)}_{\rm tadpole}+ \tilde\Phi_{\rm flower}+ \tilde\Phi_{\rm rest})-\chi_j^{(3)} \,.
\eea
A generic basketball graph is shown in Fig. (\ref{superExLABEL}a). The associated symmetry factor for a graph with two $\tilde V_j$ vertices is $(1/2)(1/j!)$. The first term on the right side of (\ref{unwrap2}) therefore gives $D^j\tilde V_j$.
Equation ({\ref{unwrap2}) can be solved iteratively to obtain an expression of the form
\bea
\label{subTilde}
\tilde V_j=D^{-j}V^c_j+f_j[D,V^c_k]\,,
\eea
 which is valid to any desired loop order.

 We define the interacting part of the effective action through the equation
\bea
\label{PhiIntdef}
\Gamma[\phi,D,V^{0}_j,V_k] =:  \Gamma^0[\phi,D,V^0_j,V_k] -i \Phi^{\rm int}[V_j]\,.
\eea
Using (\ref{dormant2point3}), we obtain
\bea
\label{can2}
\Phi^{\rm int}[V_j] = &&  \big(\tilde\Phi_{\rm basketball}+\tilde\Phi^{(2)}_{\rm tadpole}+ \tilde\Phi_{\rm flower}+ \tilde\Phi_{\rm rest}\big) -\sum_{j=3}^n \frac{1}{j!}(V_j^c + \chi^{(3)}_j) \tilde V_j
\eea
We comment that Eq. (\ref{can2}) formally expresses $\Phi^{\rm int}[V_j]$ as a functional of tilde and connected vertices. The procedure to obtain a functional of proper vertices is explained below.

The first step is to use Eq. (\ref{subTilde}) to remove the tilde vertices. We see immediately that this substitution gives
\bea
\sum_{j=3}^n \frac{1}{j!}V^c_j \tilde V_j \rightarrow 2 \sum_{j=3}^n\left[ \frac{1}{2}\frac{1}{j!}V^c_j \,D^{-j}\,V^c_j \right] +\cdots ~ =  2 \Phi^c_{\rm basketball} + \cdots
\eea
which causes the sign flip in the basketball diagrams on the right side of (\ref{can2}). We argue below that when all tilde vertices are removed the result is
\bea
\label{dormant2point4}
\Phi^{\rm int}[V_j] = -\Phi^c_{\rm basketball} + \Phi^c_{\rm flower}+ \Phi^c_{\rm rest} \,.
\eea
The function $\Phi^c_{\rm basketball}$ represents the same set of basketball diagrams as $\tilde\Phi_{\rm basketball}$ with tilde vertices replaced by connected vertices ($\tilde V_j\rightarrow D^{-j}V^c_j$). The function $\Phi^c_{\rm flower}$ contains only graphs with flower topology, but not the same set of flower graphs as $\tilde\Phi_{\rm flower}$. Similarly, $\Phi^c_{\rm rest}$ contains only graphs that are not basketball, tadpole, or flower topologies, but not the same set of graphs as $\tilde\Phi_{\rm rest}$.
Recall that the tadpole graphs are those that produce disconnected contributions to the eom that corresponds to the tadpole vertex. It is clear that if a term of the form $\Phi^c_{\rm tadpole}$ survived in Eq. (\ref{dormant2point4}), there would be a disconnected contribution to the eom for any connected vertex $V^c_j$ which is a tadpole vertex. Since disconnected terms do not appear in the perturbative expansion [from the definition of connected vertices in Eq. (\ref{Wders})], they also do not appear in the skeleton expansion.

The connected vertex can be written in terms of proper vertices $V_j$ using Eqs. (\ref{Wders}) and (\ref{properDefn}).   We will argue below that when we replace connected vertices with the appropriate expressions containing proper vertices, the flower graphs cancel and  we obtain
\bea
\label{dormant2point5}
\Phi^{\rm int}[V_j] = -\Phi_{\rm basketball} + \Phi_{\rm rest}\,.
\eea
The function $\Phi_{\rm basketball}$ contains basketball graphs which are functions of proper vertices. $\Phi_{\rm rest}$ contains only graphs that are not basketball, tadpole, or flower topologies, but they are not the same graphs as in $\Phi^c_{\rm rest}$ or $\tilde\Phi_{\rm rest}$.

Consider the form of the eom obtained by functionally differentiating the effective action with respect to the vertex $V_j$. The effective action is obtained from Eqs. (\ref{PhiIntdef}) and (\ref{dormant2point5}).
For purposes of illustration, we rewrite the result as
\bea
\Gamma[\phi,D,V^{0}_j,V_k] &=&  \Gamma^0[\phi,D,V^0_j,V_k] -i \big(-\Phi_{\rm basketball} + \Phi_{\rm rest}\big)\nonumber\,\\
=S_{cl}[\phi]&+&\frac{i}{2} {\rm Tr} \,{\rm Ln}D^{-1}  +
\frac{i}{2} {\rm Tr}\,\big[ \left(D^0\right)^{-1} D\big] -i \big(\Phi^0_{\rm basketball}+ \Phi^0_{\rm no~basketballs}-\Phi_{\rm basketball} + \Phi_{\rm rest}\big)\,,\nonumber
\eea
where we have split $\Phi^0[V^0_j,V_k]$ into two parts: $\Phi^0_{\rm basketball}$ and $\Phi^0_{\rm no~basketballs}$. In Appendix \ref{superBballSection}, we show that, in $\Phi^{\rm int}[V_j]$, the vertex $V_j$ appears in one basketball diagram, and other diagrams which are higher loop order. Similarly, in $\Phi^0[V^0_j,V_k]$ (see Appendix \ref{bareSection}) the vertex $V_j$ appears in the basketball diagram with one proper vertex and one effective bare vertex, and other diagrams which are higher loop order.
Using these results, and the fact that $\Phi_{\rm rest}$ does not contain flower diagrams, the eom for the vertex $V_j$ for $j\ge 3$ has the form
\bea
\label{EOM}
&&\frac{\delta \Gamma[\phi,D,V^{0}_j,V_k]}{\delta V_j}= 0\,,\nonumber \\
&&\Rightarrow ~~~~ j! D^{-j}\frac{\delta\Phi_{\rm basketball}}{\delta V_j}  = j! D^{-j}\frac{\delta\Phi^0_{\rm basketball}}{\delta V_j}+j! D^{-j}\frac{\delta\Phi^0_{\rm no~basketballs}}{\delta V_j}+ j! D^{-j}\frac{\delta\Phi_{\rm rest}}{\delta V_j}\,, \nonumber \\
&&\Rightarrow ~~~~ V_j  =V^0_j+{\rm fcn}^\prime_j[V^0_l,V_k]+{\rm fcn}_j[V_k]\,,
\eea
where both ${\rm fcn}^\prime_j[V_l^0,V_k]$ and ${\rm fcn}_j[V_k]$ contain only 1PI loop diagrams.
The terms in the second and third lines of Eq. (\ref{EOM}) are written in the same order.
For example, the term $V_j$ on the left side of the last equation comes from functionally differentiating the basketball diagram in $\Phi^{\rm int}$.
If the flower topologies did not cancel when the effective action is written as a function of proper vertices, there would be a 1PR contribution to the eom for any proper vertex $V_j$ which is a flower vertex.
Since 1PR terms do not appear in the perturbative expansion [from the definition of proper vertices in Eq. (\ref{properDefn})], they also do not appear in the skeleton expansion\footnote{Tadpole and flower topologies are allowed in the part of the effective action that contains bare vertices, as long as the tadpole vertices and flower vertices are bare. For example, the EIGHT diagram is a tadpole, but the 4-point vertex is bare, and thus there is no disconnected contribution to the eom for the 4-point vertex from differentiating the EIGHT graph. Similarly, the HAIR graph is a flower diagram, but again the 4-point vertex is bare, and there is no 1PR contribution to the eom for the 4-point vertex.}.

Note that,
for $j\ge 3$, there is no contribution from functionally differentiating the 1-Loop terms in $\Gamma^0[\phi,D,V^0_j,V_k]$  with respect to $V_j$. However, for $j=2$, the 1-Loop terms in $\Gamma^0[\phi,D,V^0_j,V_k]$ do contribute, and produce the terms in the eom that correspond to $V_j$ and $V_j^0$ on the left and right sides of Eq. (\ref{EOM}), respectively.

\section{Proof of Equivalence of the eom and sd equations}
\label{sdeom}

From this point on, we consider $\phi=0$ for simplicity, which means that the bare vertices $V^{oo}_j$ are equivalent to the effective bare vertices $V^{0}_j$, and the bare propagator $D^{oo}$ is equivalent to the effective bare propagator $D^0$.

In this section, we show that the eom's produced by the $n$-Loop $n$PI effective theory are equivalent to the sd equations, up to the order at which they are consistent with the underlying symmetries of the original theory\footnote{For the 2-point function, one can show without using fictitious vertices that the eom from the $n$-Loop $n$PI effective action (for any $n$) and sd equation have exactly the same form. We prove this surprising result in Appendix \ref{piSpecial}.}.
We comment that although both the $n$PI eom's and the sd  equations  are sets of coupled nonlinear integral equations that contain nonperturbative physics, there are
significant differences between them.
For an $n$PI effective theory, the effective action is truncated, and the resulting eom's form a closed set. In contrast, the sd equations form an infinite hierarchy of coupled equations which must be truncated in order to do calculations.
In addition, there are fundamental differences in the basic structure of the two sets of equations. In the sd equation, all graphs contain one bare vertex and are not symmetric with respect to permutations of external legs. The $n$PI eom's are symmetric and (for $n>2$) some graphs contain no bare vertices.

The first step is to compare the perturbative expansions of the sd equations and the eom's. In order to do this, we must use the equations of motion in (\ref{EOM}), and also the corresponding equation for the 2-point function which is obtained from $\delta\Gamma/\delta D = 0$. The complete set of equations can be written
\bea
\label{EOM2}
&& V_j = V_j^0 + {\rm fcn}^\prime_j[V_l^0,V_k]+ {\rm fcn}_j[V_k]\,,~~\{j,k,l\}\ge 2\,.
\eea
The definitions of the functions ${\rm fcn}^\prime_j[V_l^0,V_k]$ and ${\rm fcn}_j[V_k]$ are given in Eq. (\ref{EOM}) for $j\ge 3$. For $j=2$, we have
\bea
\label{EOMforce}
j=2:~~&&{\rm fcn}^\prime_2[V_l^0,V_k] =  - 2 \frac{\delta \Phi^0[V_l^0,V_k]}{\delta D}\,,~~~~{\rm fcn}_2[V_k] =  - 2\frac{\delta  \Phi^{\rm int}[V_k]}{\delta D}\,.
\eea
If we define both terms in the 1-loop effective action to be 1-loop basketballs, the terms $V_j$ and $V_j^0$ in (\ref{EOM2}) come from the functional derivative acting on the $(j-1)$-loop basketball graph, for each value of $j$.
The sign difference for the 2-point function and the missing factor $D^{-2}$ occurs because of the fact that it is conventional to write the effective action as a function of the propagator $D$ instead of the inverse propagator $D^{-1}$ (see footnote 2).
To illustrate the notation, we write out Eq. (\ref{EOM2}) for $j=2$ and $j=3$:
\bea
\label{2ptEx}
\frac{\delta \Gamma[V_k]}{\delta D}=0~~\to~~D^{-1} && = (D^0)^{-1}-2\frac{\delta \Phi^0[V_l^0,V_k]}{\delta D} -2\frac{\delta \Phi^{\rm int}[V_k]}{\delta D} \\[2mm]
&& =: (D^0)^{-1}-\Pi[V_k]\,,\nonumber\\[2mm]
 \frac{\delta \Gamma[V_k]}{\delta V_3}=0~~\to~~V_3 && = V_3^0 +3!D^{-3}~\frac{\delta \Phi_{\rm no~basketballs}^0[V_l^0,V_k]}{\delta V_3} +3!D^{-3}~\frac{\delta \Phi_{\rm rest}}{\delta V_3} \nonumber \\[2mm]
&& = V_3^0+ {\rm fcn}_3^\prime[V_l^0,V_k] + {\rm fcn}_3[V_k]\,.\nonumber
\eea

 We can generate the perturbative expansion of any functional of proper vertices by repeatedly substituting (\ref{EOM2}). We can also repackage a perturbative set of diagrams as skeleton diagrams that contain proper vertices by repeatedly using the same equation in the form
\bea
\label{subbereom}
&&V^0_j= V_j-{\rm fcn}^\prime_j[V_l^0,V_k]-{\rm fcn}_j[V_k]\,,~~\{j,k,l\}\ge 2\,.
\eea
In the rest of this section, we use $\{j,k,l\}\ge 2$.

\begin{description}
\item[item 1] If we convert a set of skeleton diagram for the vertex $V_j$ into a series of perturbative diagrams using (\ref{EOM2}), the leading loop order of the new set of diagrams is greater than or equal to the leading loop order of the original set.
\item[item 2] If we include fictitious vertices $V^0_j$ for $5\le j \le n$, we can convert skeleton diagrams to perturbative diagrams using (\ref{EOM2}), or perturbative diagrams to skeleton diagrams using (\ref{subbereom}), and the leading loop order of the new set of diagrams is equal to the leading loop order of the original set.
\end{description}

We illustrate these statements with an example. We use $L_{pt}$ to indicate the loop order of the perturbative expansion.
Consider the skeleton diagram shown in part $(a)$ of Fig. \ref{ptExample}, which is of order $L=2$. We can expand this diagram as a series of perturbative diagrams using equations of the form (\ref{EOM2})  which are shown for this example in part $(b)$ of the figure\footnote{The propagators in the skeleton diagrams in Fig. \ref{ptExample} also have to be expanded to obtain a perturbative diagram. This will produce extra loops that correspond to self-energy corrections. In this paper, we do not introduce notation to distinguish skeleton and perturbative propagators in diagrams.}. The leading order term is shown in part $(c)$, and is of order $L_{pt}=2$. Thus, we have $L=L_{pt}=2$. Now, consider the result if we set $V_5^0=0$, which means we remove the first diagram on the right side of part $(b_2)$. In this case, the leading order term is shown in part $(d)$ and is of order $L_{pt}=3$. Thus, we see that if the fictitious vertex $V_5^0$ is set to zero we have $L_{pt} > L$.
\par\begin{figure}[H]
\begin{center}
\includegraphics[width=12cm]{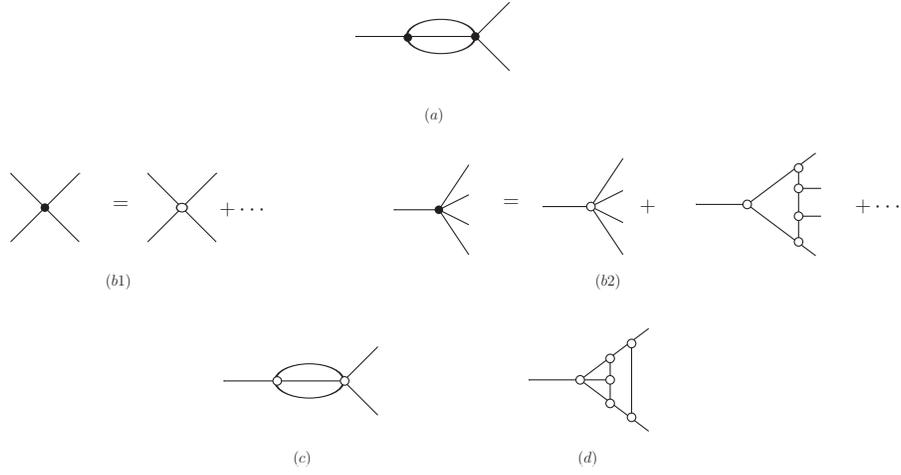}
\end{center}
\caption{\label{ptExample} Diagrams used to explain items 1 and 2.}
\end{figure}

We consider truncating the $n$PI effective action at $m$-loop order\footnote{At $m$-Loops, the $n$PI effective action is the same as the $(n+1)$PI effective action for $n\ge m$ \cite{berges}.}. The functional derivative of an $m$-loop graph with respect to the variational vertex $V_j$ opens $j-1$ loops. This means that an arbitrary $m$-loop graph in the effective action which contains the vertex $V_j$ produces a term with ${\cal L}[m,j]$ loops in the skeleton expansion of the eom for the vertex $V_j$, where we  define
\bea
\label{calL}
{\cal L}[m,j] := m-j+1\,.
\eea
Note that the order of the original $m$-loop graph in the effective action corresponds to $j=1$.

Now, we consider the effect of adding an arbitrary $(m+1)$ loop graph to the skeleton expansion of the $m$-Loop $n$PI effective action.
This $(m+1)$ loop graph will produce new contributions to the skeleton expansions of the eom's for each vertex contained in the graph. There are two kinds of contributions:

\vspace{.2cm}

\noindent (1) Taking the functional derivative of this added graph with respect to $V_{ j}$ produces new terms in the skeleton expansions of the $V_{j}$ eom of order ${\cal L}[m+1,j]$. These new terms contribute at $L={\cal L}[m+1,j]$ loops in the skeleton expansion and (using item 1) $L_{pt}\ge {\cal L}[m+1,j]$ loops in the perturbative expansion.

\vspace{.2cm}

\noindent (2) We also need to consider lower loop diagrams in the skeleton expansion of $V_{j}$ of order ${\cal L}[m^\prime,j]$ ($m^\prime\le m$), with an arbitrary variational vertex $V_k$ replaced by a term in its eom  which was produced by functional differentiation of the $(m+1)$ loop graph that was added to the effective action. For any $k$, the new contributions to the vertex $V_k$ from this added graph are of order ${\cal L}[m+1,k]$. The substitution of vertex $V_k$ produces terms of order $L={\cal L}[m^\prime,j]+{\cal L}[m+1,k]$ in the skeleton expansion of $V_{j}$. Using Eq. (\ref{calL}) and $k_{\rm max}=m^\prime +1$ (see Appendix \ref{superBballSection}), we obtain  $L\ge {\cal L}[m^\prime,j]+{\cal L}[m+1,k_{\rm max}]= {\cal L}[m+1,j]$. Thus, we have shown that these terms also contribute to the eom of the vertex $V_j$ at $L={\cal L}[m+1,j]$ loops in the skeleton expansion and (using item 1) $L_{pt}\ge {\cal L}[m+1,j]$ loops in the perturbative expansion.

\vspace{.2cm}

We conclude that if we add an arbitrary $(m+1)$ loop graph to the skeleton expansion of the $m$-loop $n$PI effective action, this graph will produce terms at ${\cal L}[m+1,j]$ loops or higher in both the skeleton and perturbative expansions of the eom for the vertex $V_{j}$.
Furthermore, we know that without truncation, the expanded effective action and equations of motion for the vertices $V_{j}$ must exactly match the 1PI perturbative expansion.  The conclusion is
\begin{description}
\item[item 3] The $m$-Loop $n$PI effective action produces all terms in the perturbative expansions of the effective action and the equations of motion for the vertices $V_{j}$ up to $L_{pt}= {\cal L}[m,j]$ loops.
\end{description}
Equivalently, the  vertex functions have the correct crossing symmetry to $L_{pt}={\cal L}[m,j]$ loops.  We say that the variational vertex functions respect crossing symmetry to the ``truncation order.''

\ts

Now we consider the Schwinger-Dyson equations, which form an infinite hierarchy of coupled nonlinear integral equations. They have the form
\bea
\label{sdGeneral}
V_j^{sd} = V_j^0+{\rm fcn}_j^{sd}[V_l^{0},V_k^{sd}]\,.
\eea
Although the structure of the sd equations is very different from the eom, when we truncate the sd equations by setting $V^{sd}_{m+k}=V^{0}_{m+k}$ for $k\ge 1$, the vertex $V^{sd}_{j}$ also matches the perturbative expansion up to $L_{pt}= {\cal L}[m, j]$ loops \cite{paper1}.

If we truncate at some given number of loops, the most general effective action is obtained by considering the same number of variational vertices (see footnote 7). For this reason, from this point on, we consider only $m=n$. Using item 3, we have that

\begin{description}
\item[item 4] The perturbative expansions of the $n$-loop $n$PI eom's and the sd equations truncated by setting $V^{sd}_{n+k}=V^{0}_{n+k}$ for $k\ge 1$ both match the perturbative expansion obtained from the 1PI effective action, and therefore each other, to order $L_{pt}= {\cal L}[n,j]$.
\end{description}

We can formally write Eq. (\ref{sdGeneral}) as
\bea
\label{newsd}
&& V_j^{sd} = V_j^0+{\rm fcn}^{\prime}_j[V_l^0,V_k^{sd}]+ I_j[V_l^0,V_k^{sd}]\,,\\[2mm]
&& I_j[V_l^0,V_k^{sd}]:= {\rm fcn}_j^{sd}[V_l^0,V_k^{sd}]-{\rm fcn}^{\prime}_j[V_l^0,V_k^{sd}]\,.\nonumber
\eea
We can rewrite $I_j[V_l^0,V_k^{sd}]$ as
\bea
\label{sub}
I_j[V_l^0,V_k^{sd}]={\rm fcn}_j[V_k^{sd}]~~+ ~~{\rm extra}\,,
\eea
where the functional ${\rm fcn}_j[V_k^{sd}]$ can be taken to be the same functional as in (\ref{EOM2}), since the extra term is defined to absorb any leftovers.
Comparing (\ref{EOM2}) and (\ref{newsd}) and using item 4, it is clear that ${\rm fcn}_j[V_k]$ and $I_j[V_l^0,V_k^{sd}]$ must match each other in the perturbative expansion to order $L_{pt}={\cal L}[n,j]$.Therefore, we know that the extra term is of order $L_{pt}={\cal L}[n,j]+1$.
Using item 2, the extra term can be rewritten as a series of skeleton diagrams of order $L={\cal L}[n,j]+1$. Thus, we have shown
\begin{description}
\item[item 5] The  sd equations can be rearranged to have the same form as the $n$PI eom's, plus additional terms of order $L= {\cal L}[n,j]+1$ in the skeleton expansion.
\end{description}

\ts

In the next section, we use the result in item 5 to calculate the effective action, without taking a Legendre transform. We emphasize that the proof of this result depends on the use of fictitious bare vertices $V_j^{0}$ for $5\le j \le n$. Specifically, items 1, 3, and 4 are true with or without fictitious vertices, but items 2 and 5 are only true when these vertices are included.

\section{A new approach to the calculation of the effective action}
\label{newApproachSection}

In this section, we explain the technique of a new approach to calculate (the interacting part of) the $n$-Loop $n$PI effective action. The basic idea is to calculate the sd equations, using standard techniques (see Refs. \cite{cvitanovic,kajantie,alkoferALG}), and then exploit the fact that they can be re-arranged to have the same form as the $n$PI equations of motion, up to the truncation order (see Sec. \ref{sdeom}). One joins the legs of each graph in the rearranged sd equations, to get the structure of the graphs in the effective action. It is clear that this procedure will produce all of the graphs in the effective action, at a given order. The trick is to obtain the correct symmetry factor. In order to see how this can be done, consider starting from a known result for the effective action $\Phi^{\rm int}$, taking derivatives of each graph with respect to each variational vertex, and trying to reconstruct the effective action by joining the legs in each of these eom's. There are two potential difficulties:

(1)
A given graph in the effective action will produce contributions to the eom's of each vertex it contains. In order to produce the correct symmetry factor when joining legs, we must drop the corresponding contribution in all but one eom, which we take to be the eom for the largest vertex present. For example, the TARGET graph (see Fig. \ref{fig:PHIint2}) gives a contribution to the eom of the vertices $V_3$ and $V_4$. We drop the contribution to the $V_3$ eom and recover the TARGET graph by joining the legs of the contribution to the $V_4$ eom. If we did not drop contributions to the $V_3$ eom from the TARGET graph, we would produce unwanted copies of the TARGET graph when we joined the legs in the graphs in the full $V_3$ eom.

(2) If the largest vertex in a given diagram in the effective action appears more than once, the graph that is produced by joining the legs will have a symmetry factor that is too large by a factor equal to the number of times the vertex appears.
For example, consider the LOOPY graph (see Fig. \ref{fig:PHIint2}). The largest vertex is $V_4$ which appears 3 times. If we join the legs on the contributions of the LOOPY graph to the $V_4$ eom, we recover the LOOPY graph, but with a symmetry factor  which is 3 times too big.

For any diagram in the equation of motion for the largest vertex $V_j$, the correct symmetry factor for the contribution to the effective action that we get by joining legs is
\bea
\label{calS}
S[j]=s(1/v_j)(1/j!)\,,
\eea
where $s$ is the numerical factor in front of the diagram in the eom, and $v_j-1$ is the number of times the vertex $V_j$ appears in this diagram.

We give two examples of how to use this formula. In the TARGET graph, the largest vertex is $V_4$ which appears only once, and there are three contributions to the eom for the vertex $V_4$, all of which have $s=1$ and $v_4-1=0$. Joining legs and using (\ref{calS}) we recover the TARGET graph with symmetry factor $3\cdot (1)\cdot (1/4!)=1/8$ (see Fig. \ref{fig:PHIint2}).
In the LOOPY diagram, there are 3 contributions to the eom for the vertex $V_4$ (which correspond to the $s$, $t$, and $u$ channels), all of which have $s=1/2$ and $v_4-1=2$. Joining legs and using (\ref{calS}), we recover the LOOPY graph with symmetry factor $3/2\cdot (1/3)\cdot (1/4!) = 1/48$ (see Fig. \ref{fig:PHIint2}).

The complete set of rules to generate the interacting part of the $n$-Loop $n$PI effective action from the sd equations is given below. In Secs. \ref{example44} and \ref{example55}, we describe in detail how the procedure works for the 4-loop 4PI effective action and the 5-loop 5PI effective action.

\begin{description}

\item[Step 1] In the classical action, include bare vertices $V_j^{0}$ for $3\le j\le n$.

\item[Step 2] Using this new classical action, derive $\Phi^0$ and the sd equations for the vertices $V_j$ for $3\le j \le n$, using standard techniques.

\item[Step 3] Extract from $\Phi^0$ the functions ${\rm fcn}^{\prime}_j[V_l^0,V_k^{sd}]$ which are defined in (\ref{EOM}). Once these functions have been obtained, set the fictitious vertices to zero in $\Phi^0$ (and thus remove the nonrenormalizable interactions). The resulting expression for $\Phi^0$ is the same for all $n$PI effective actions with $n\ge 4$.

\item[Step 4]
Rearrange (\ref{sdGeneral}) in the form (from this point on, we suppress the superscript ``$sd$'' on vertices):
\bea
\label{subbersd}
&& V^0_j = V_j-{\rm fcn}^{sd}_j[V_l^0,V_k]\,.
\eea
Following the procedure described below, use (\ref{subbersd}) to remove the bare vertices in the functionals $I_j[V_l^{0},V_k]$ defined in (\ref{newsd}) for $j=n,n-1,\cdots, 3.$

\begin{description}

\item[Level 1] Use (\ref{subbersd}) to remove bare vertices in $I_n$ until all terms with ${\cal L}[n,n] := 1$ loop contain no bare vertices. Drop terms in $I_n$ with more than ${\cal L}[n,n]$ loops and terms containing $V_k$ with $k\ge n+1$ (because these vertices do not contribute to the $n$PI effective action). Join the legs of the remaining terms and calculate the symmetry factor $S[n]$ using (\ref{calS}).

\item[Level 2] Use (\ref{subbersd}) to remove bare vertices in $I_{n-1}$ until all terms with ${\cal L}[n,n-1] := 2$ Loops contain no bare vertices. Drop terms  in $I_{n-1}$ with more than ${\cal L}[n,n-1]$ loops and terms containing $V_k$ with $k\ge n$. Join the legs of the remaining terms and calculate the symmetry factor $S[n-1]$ using (\ref{calS}).

$\vdots$

\item[Level j] Use (\ref{subbersd}) to remove bare vertices in $I_{n-j+1}$ until all terms with ${\cal L}[n,n-j+1] $ Loops contain no bare vertices. Drop terms in $I_{n-j+1}$ with more than ${\cal L}[n,n-j+1]$ loops and terms containing $V_k$ with $k\ge n-j+2$. Join the legs of the remaining terms and calculate the symmetry factor $S[n-j+1]$ using (\ref{calS}).

$\vdots$

\item[Level n-2] Use (\ref{subbersd}) to remove bare vertices in $I_3$ until all terms with ${\cal L}[n,3] := n-2$ Loops contain no bare vertices. Drop terms in $I_3$ with more than ${\cal L}[n,3]$ loops and terms containing $V_k$ with $k\ge 4$. Join the legs of the remaining terms and calculate the symmetry factor $S[3]$ using (\ref{calS}).

\end{description}

\item[Step 5] Add the basketball diagrams with two proper vertices $V_{j}$ for $3 \le j \le n$. The symmetry factor for each graph is $-1/(2 (j)!)$.

\end{description}

\section{Example of 4-Loop 4PI effective action.}
\label{example44}

In this section, we calculate the 4-Loop 4PI effective action and verify that our technique produces the known result \cite{EK4}, which is reproduced in Figs. \ref{fig:Phi0} and \ref{fig:PHIint2} for convenience.
\par\begin{figure}[H]
\begin{center}
\includegraphics[width=10cm]{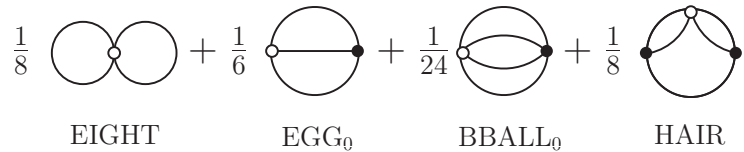}
\end{center}
\caption{\label{fig:Phi0}$\Phi^0$ for a theory with bare vertices $V_3^{0}$ and $V_4^{0}$. }
\end{figure}
\begin{figure}[H]
\begin{center}
\includegraphics[width=15cm]{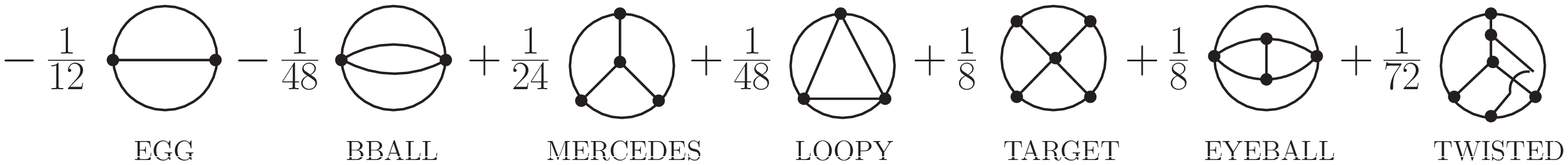}
\end{center}
\caption{\small 4-Loop diagrams contributing
to $\Phi^{\rm int}$.} \label{fig:PHIint2}
\end{figure}

We follow the procedure outlined in the previous section.

\noindent {\bf Step 1:} Start with a classical action that includes bare 3-point and 4-point vertices: $V_3^{0}$ and $V_4^{0}$.

\noindent {\bf Step 2:} Calculate $\Phi^0$ and the sd equations using this action. The sd equations for $V_3$ and $V_4$ are reproduced from \cite{paper1} in Figs. \ref{SD3} and \ref{SD4}.
\par\begin{figure}[H]
\begin{center}
\includegraphics[width=14cm]{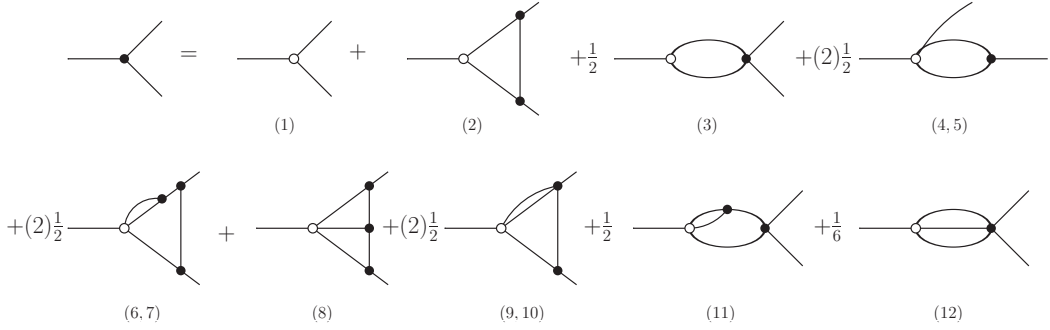}
\end{center}
\caption{\label{SD3}Schwinger-Dyson equation for the 3-point vertex with $V_3^0$ and $V_4^0$.}
\end{figure}
\par\begin{figure}[H]
\begin{center}
\includegraphics[width=15cm]{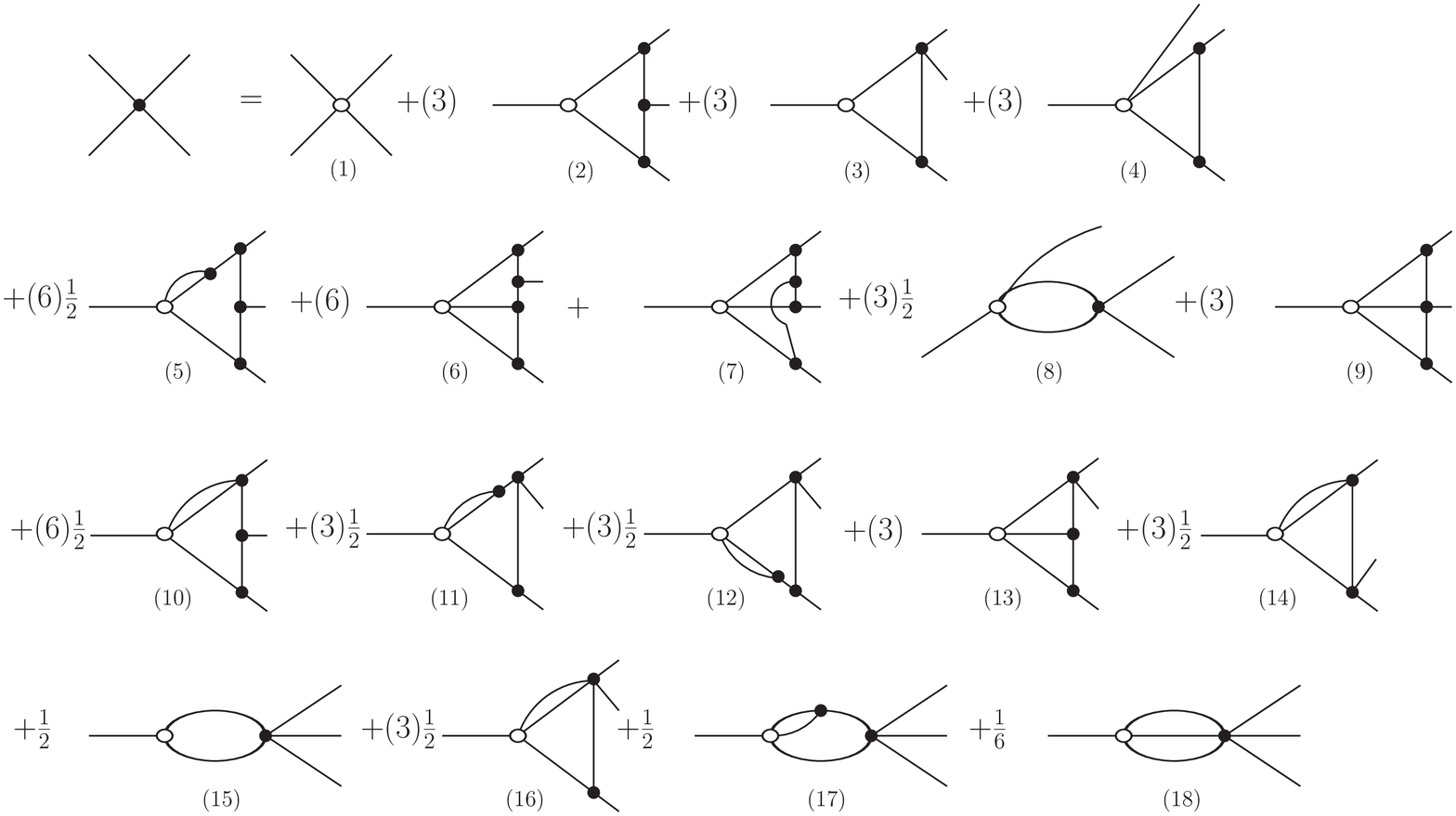}
\end{center}
\caption{\label{SD4}Schwinger-Dyson equation for the 4-point vertex with $V_3^0$ and $V_4^0$.}
\end{figure}

\noindent {\bf Step 3:} Extract ${\rm fcn}^{\prime}_3[V_l^0,V_k]$ and ${\rm fcn}^{\prime}_4[V_l^0,V_k]$. The result for ${\rm fcn}^{\prime}_3[V_l^0,V_k]$ is shown in Fig. \ref{goal34} and ${\rm fcn}^{\prime}_4[V_l^0,V_k]=0$.
\par\begin{figure}[H]
\begin{center}
\includegraphics[width=6cm]{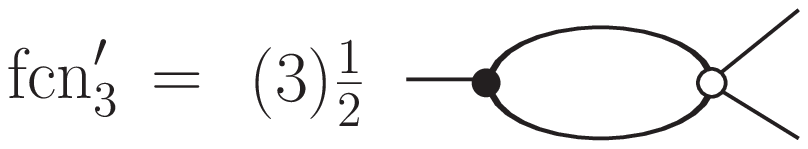}
\end{center}
\caption{\label{goal34} The result for ${\rm fcn}^{\prime}_3[V_l^0,V_k]$. Joining the legs produces the HAIR graph, and calculating the symmetry factor using Eq. (\ref{calS}) gives $(3/2)(1/2)1/3!=1/8$, which agrees with Fig. \ref{fig:Phi0}.}
\end{figure}

\noindent {\bf Step 4:}

\noindent {\bf Level 1:} We want to obtain a 1-Loop expression for $I_4$ that does not contain bare vertices. Since fcn$_4^\prime=0$, we simply take the 1-loop diagrams in Fig. \ref{SD4} and set $V^0_j=V_j$ and $V_{j\ge 5}=0$. The three graphs that are produced are shown in Fig. \ref{1loopv4}. The numbers in brackets under each diagram indicate the corresponding diagrams in the sd equation. For example, the first diagram in Fig. \ref{1loopv4} comes from the graph marked (2) in Fig. \ref{SD4} with $V_3^0=V_3$. Joining legs, the three diagrams in Fig. \ref{1loopv4} produce, respectively, the TARGET, EYEBALL, and LOOPY topologies. We calculate the symmetry factors for each graph using (\ref{calS}). For the three graphs in Fig. \ref{1loopv4} we have $v_4=1$, $v_4=2$, and $v_4=3$. The symmetry factors are 3(1)(1/4!) = 1/8,  6(1/2)(1/4!) = 1/8, and 3/2(1/3)(1/4!)=1/48, which reproduces the result in Fig. \ref{fig:PHIint2} for the TARGET, EYEBALL, and LOOPY graphs.
\par\begin{figure}[H]
\begin{center}
\includegraphics[width=12cm]{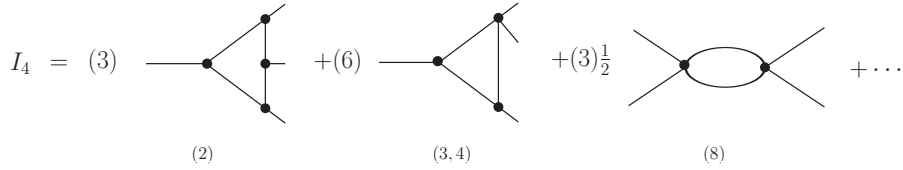}
\end{center}
\caption{\label{1loopv4} The 1-loop terms in $I_4$ with bare vertices removed. The numbers in brackets under each diagram indicate the corresponding diagrams in the sd equation (see Fig. \ref{SD4}).}
\end{figure}

\noindent {\bf Level 2:}
We want to obtain an expression for $I_3$ that does not contain bare vertices at the 2-Loop level.
We start with the 2-Loop diagrams in Fig. \ref{SD3} and subtract ${\rm fcn}_3^\prime$ (see Fig. \ref{goal34}). We set $V_{j\ge 4}=0$ (recall that vertices with $j\ge 5$ are set to zero because they are not part of the 4PI effective action, and vertices $V_4$ are set to zero to avoid double counting contributions that were obtained in Level 1 above).  This produces graphs that we refer to as ``explicit 1-loop'' and ``explicit 2-loop''. After the bare vertices are removed using (\ref{subbersd}), the explicit 1-loop graphs will produce 2-loop contributions.

\noindent \underline{explicit 2-loop}: We remove bare vertices in the explicit 2-loop graphs by setting $V_j^0=V_j$. Using $V_{j\ge 4}=0$, all 2-loop graphs drop out.

\noindent \underline{explicit 1-loop}:  The explicit 1-loop contributions to $I_3$ with $V_{j\ge 4}=0$ are shown in Fig. \ref{fig:11part1}.  We remove bare vertices using (\ref{subbersd}), iterated so that there are no bare vertices in the 1-Loop terms. These iterated expressions with $V_{j\ge 4}=0$ are shown in Fig. \ref{fig:11part2}. The result of substituting Fig. \ref{fig:11part2} into Fig. \ref{fig:11part1} is shown in Fig. \ref{fig12}. Note that the 2 graphs which would produce a 4PR contribution to the effective action cancel identically.
The final step is to join the legs and calculate the symmetry factor using (\ref{calS}). The surviving diagram from the top line in Fig. \ref{fig12} produces the MERCEDES graph, and the survivor from the second line produces the TWISTED graph (see Fig. \ref{fig:PHIint2}).
\par\begin{figure}[H]
\begin{center}
\includegraphics[width=8cm]{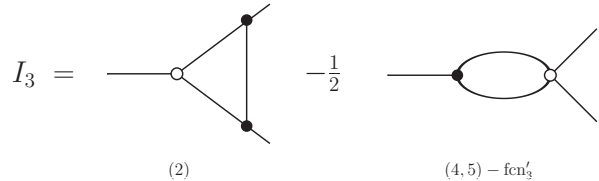}
\end{center}
\caption{\label{fig:11part1} The 1-loop terms in $I_3$. The numbers in brackets under each diagram indicate the corresponding graph in the sd equation in Fig. \ref{SD3}, and fcn$^\prime_3$ is shown in Fig. \ref{goal34}. }
\end{figure}
\par\begin{figure}[H]
\begin{center}
\includegraphics[width=8cm]{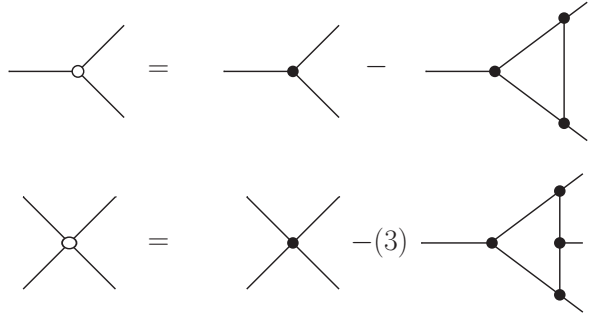}
\end{center}
\caption{\label{fig:11part2} The result of iterating Eq. (\ref{subbersd}) so that 1-loop graphs do not contain bare vertices.}
\end{figure}
\par\begin{figure}[H]
\begin{center}
\includegraphics[width=10cm]{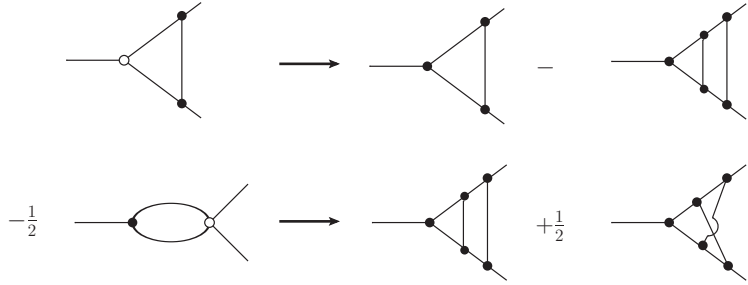}
\end{center}
\caption{\label{fig12} The diagram obtained by substituting Fig. \ref{fig:11part2} into Fig. \ref{fig:11part1}.}
\end{figure}

\noindent {\bf Step 5:} Add the basketball diagrams  which are the EGG and BBALL diagrams (see Fig. \ref{fig:PHIint2}).

\section{5-Loop 5PI effective action.}
\label{example55}

For $n\ge 5$, we need to introduce fictitious bare vertices. We illustrate the role of these vertices by describing the 5-loop 5PI calculation. Some details are left to Appendix \ref{5PIapp}.
The result of the calculation is known \cite{paper1}. For convenience, we reproduce in Fig. \ref{fig:fig17} the 5-loop diagrams.
\begin{figure}[H]
\begin{center}
\includegraphics[width=15cm]{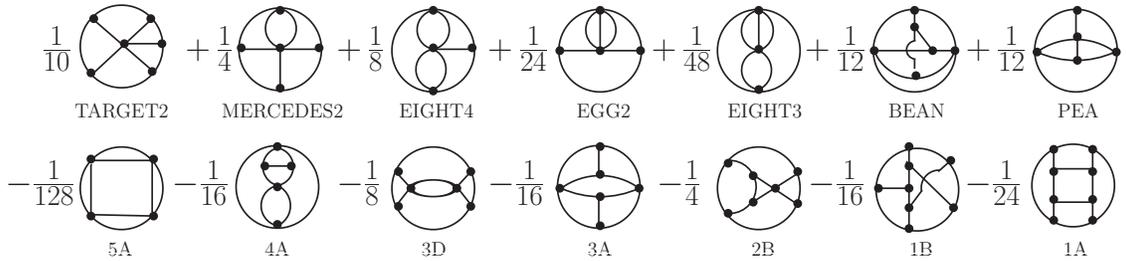}
\end{center}
\caption{\small 5-loop diagrams contributing
to $\Phi^{\rm int}$ for the 5PI effective action.} \label{fig:fig17}
\end{figure}
We follow the steps in Sec. \ref{newApproachSection}.

\noindent {\bf Step 1:} We include bare vertices $V_3^{0}$, $V_4^{0}$, and $V_5^{0}$ in the Lagrangian.

\noindent {\bf Step 2:} We calculate $\Phi^0$ and the sd equations for the vertices $V_j$, $3\le j\le 5$. The result for $\Phi^0$ is given in Fig. \ref{fig:Phi0} and the first line of Fig. \ref{fig:Phi0new}, excluding the basketball diagram with the vertex $V_6^{0}$.

The fictitious vertices produce many contributions to the sd equations, but not all are needed to calculate the 5-loop 5PI effective action.

In Step 4, we will need to calculate $I_5$ to 1-Loop level, which means we only need 1-Loop terms in the sd equation for $V_5$. These 1-loop diagrams are the first 10 graphs on the right side of Fig. \ref{SD5}.
\par\begin{figure}[H]
\begin{center}
\includegraphics[width=15cm]{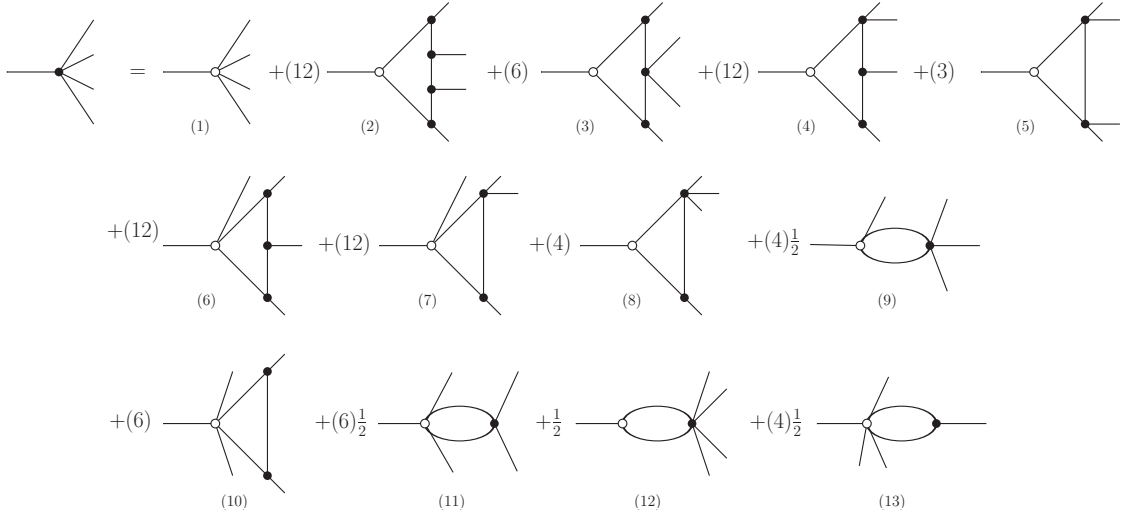}
\end{center}
\caption{\label{SD5}1-Loop terms in the Schwinger-Dyson equation for the 5-point vertex with $V_j^0$ for $3\le j\le 6$.}
\end{figure}
In addition, the sd equations for the vertices $V_3$ and $V_4$ in Figs. \ref{SD3} and \ref{SD4} receive extra contributions from diagrams with the vertex $V_5^0$. For $V_4$, we need 1-loop diagrams with $V_{j\ge 5}=0$ and 2-loop diagrams with $V_j^0=V_j$ and $V_{j\ge 5}=0$. For $V_3$, we need 2-Loop diagrams with $V_{j\ge 4}=0$ and 3-loop diagrams with $V_j^0=V_j$ and $V_{j\ge 4}=0$. These extra diagrams are shown in Fig. \ref{extrasd}.
\par\begin{figure}[H]
\begin{center}
\includegraphics[width=11cm]{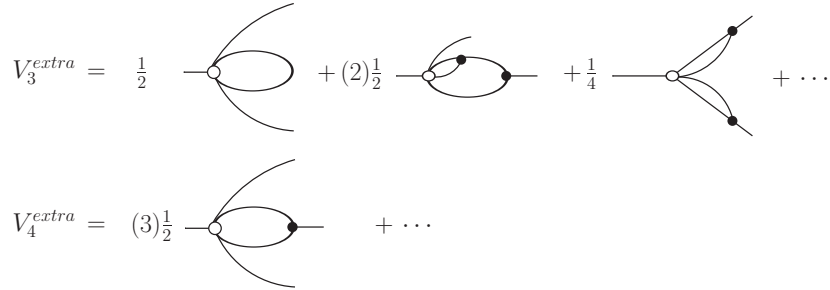}
\end{center}
\caption{\label{extrasd} Some extra terms in the sd equations for $V_3$ and $V_4$ containing $V_5^0$.}
\end{figure}

\noindent {\bf Step 3:} From $\Phi^0$, we extract ${\rm fcn}^{\prime}_i$ with $i=3 \, {\rm and}\, 4$. The results are shown in  Fig. \ref{extra}. We have ${\rm fcn}_5^{\prime}=0$ at the 5-Loop 5PI level.

\par\begin{figure}[H]
\begin{center}
\includegraphics[width=16cm]{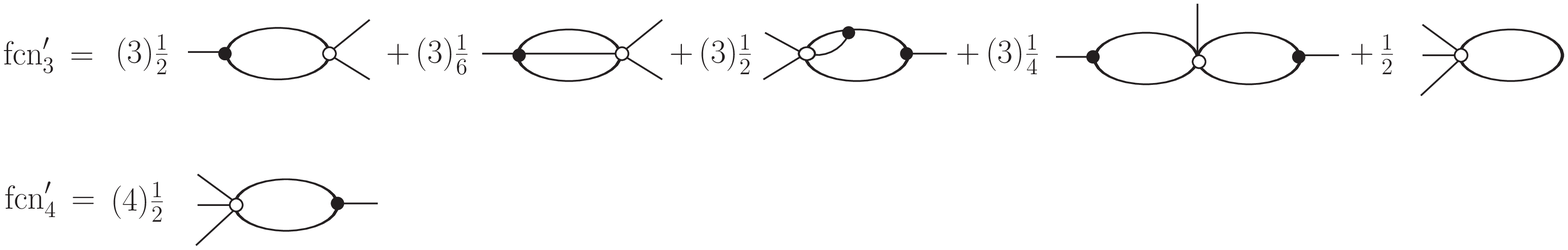}
\end{center}
\caption{\label{extra} Results for ${\rm fcn}^{\prime}_3$ and ${\rm fcn}^{\prime}_4$.}
\end{figure}

\noindent {\bf Step 4:}

\noindent {\bf Level 1:} We construct $I_5$ at the 1-loop level.
The 1-loop terms in the sd equation are shown in Fig. \ref{SD5}, and ${\rm fcn}_5^{\prime}=0$. We replace bare vertices with proper ones $V_j^0=V_j$ and set $V_{j\ge 6}=0$, which removes the last two diagrams. The surviving terms are shown in Fig. \ref{fig:15part1}.
\par\begin{figure}[H]
\begin{center}
\includegraphics[width=17cm]{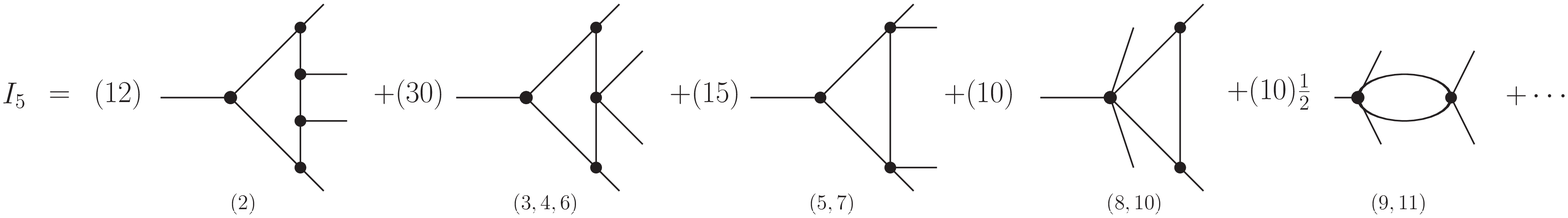}
\end{center}
\caption{\label{fig:15part1} One-loop contributions to $I_5$. The numbers in brackets under each diagram indicate the corresponding graph in the sd equation in Fig. \ref{SD5}.}
\end{figure}
The final step is to join the legs and calculate the symmetry factor from (\ref{calS}).  The fourth graph has $v_5=2$ and $S=10\cdot(1/2)\cdot(1/5!)=1/24$, which reproduces EGG2, and the fifth has $v_5=2$ and $S=5\cdot(1/2)\cdot(1/5!)=1/48$, which reproduces EIGHT3 (see Fig. \ref{fig:fig17}).
Note that the graphs marked (10) and (11) in Fig. \ref{SD5}, which contain the vertex $V_5^0$, are needed to obtain these results. This is an example of the role of the fictitious bare vertices. The first three graphs in Fig. \ref{fig:15part1} produce the TARGET2, MERCEDES2, and EIGHT4 graphs, respectively, (see Fig. \ref{fig:fig17}).

\noindent {\bf Levels 2 and 3:} We need to construct $I_4$ at the 2-Loop level and $I_3$ at the 3-Loop level. Some details of the calculation are given in Appendix \ref{5PIapp}.

\noindent {\bf Step 5:} We add the 2-, 3-, and 4-loop basketballs.\\

Combining all pieces, we reproduce the 5-Loop 5PI effective action, which was obtained through a much more lengthy calculation in \cite{paper1}, using Legendre transforms.

\section{Result for 6-Loop 6PI}
\label{example66}

The 6-Loop 6PI effective action can be calculated using the same method.

\noindent {\bf Step 1:} We include bare vertices $V_j^0$ for $3\le j\le 6$ in the Lagrangian.

\noindent {\bf Step 2:} The additional terms in $\Phi^0$ and the sd equation for the self-energy which contain $V_5^0$ and $V_6^0$ are shown in Figs. \ref{fig:Phi0new} and \ref{sdnewFIG}, respectively.
\par\begin{figure}[H]
\begin{center}
\includegraphics[width=14cm]{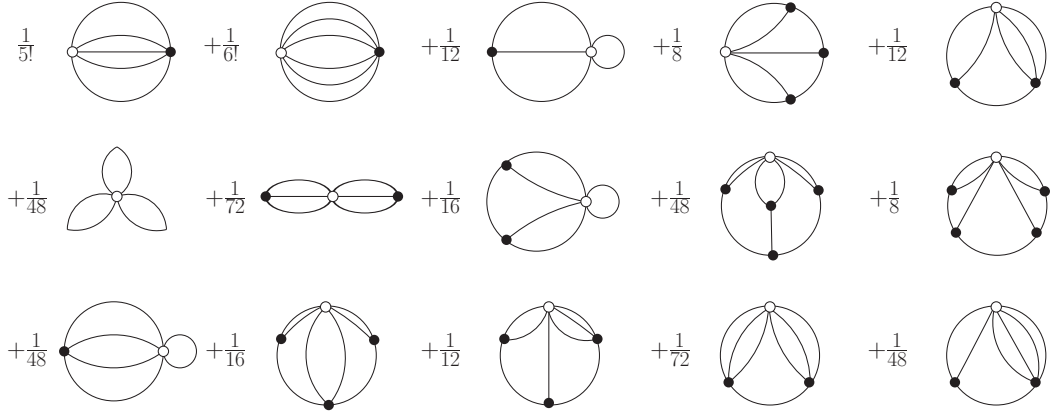}
\end{center}
\caption{\label{fig:Phi0new} 5-Loop contributions to $\Phi^0$ that contain the bare vertices $V_5^0$ and $V_6^0$.}
\end{figure}
\par\begin{figure}[H]
\begin{center}
\includegraphics[width=10cm]{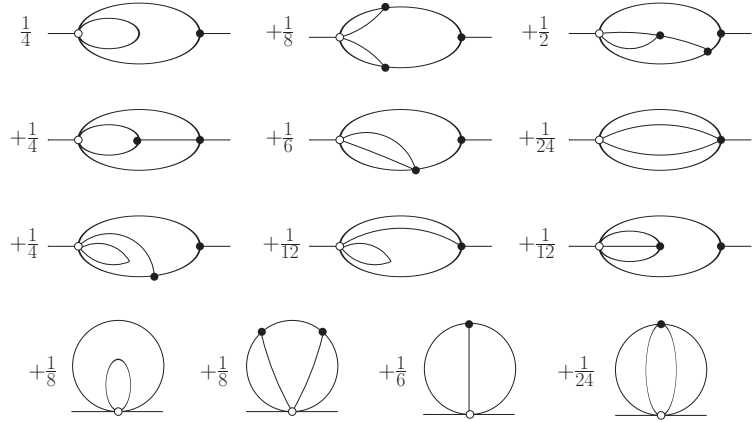}
\end{center}
\caption{\label{sdnewFIG}3-Loop contributions to the sd equation for $\Pi$ from terms with $V_5^0$ and $V_6^0$. }
\end{figure}

Using these expressions, it is straightforward to generate the corresponding results for the sd equations for $V_j$ for $3\le j\le 6$. After combining permutations of external indices, the sd equation for the vertex $V_6$ contains 20 1-loop terms, the equation for $V_5$ contains  12 1-loop terms, and 62 2-loop terms, the equation for $V_4$ contains 7 1-loop terms, 27 2-loop terms, and 88 3-loop terms, and the equation for $V_3$ contains 5 1-loop terms, 12 2-loop terms, 31 3-loop terms, and 49 4-loop terms\footnote{The 3-loop contributions to the $V_4$ equation are not needed since they all contain $V_5^0$ or $V_6^0$ and thus drop out when we set $V_j^0=V_j$ and $V_j=0$ for $j\ge 5$. Similarly, the 4-loop contributions to the $V_3$ equation are not needed since they all contain $V_6^0$ and drop out when we set $V_j^0=V_j$ and $V_j=0$ for $j\ge 4$.}.

It is straightforward to follow the procedure outlined in Sec. \ref{newApproachSection} and illustrated in Secs. \ref{example44} and \ref{example55}. The calculation can be done using Mathematica. We give only the result below.
Diagrams with highest $V_j$ equal to $V_6$, $V_5$, $V_4$, and $V_3$ are shown in Figs. \ref{aa}, \ref{bb}, \ref{cc} and \ref{dd}, respectively.
\par\begin{figure}[H]
\begin{center}
\includegraphics[width=14cm]{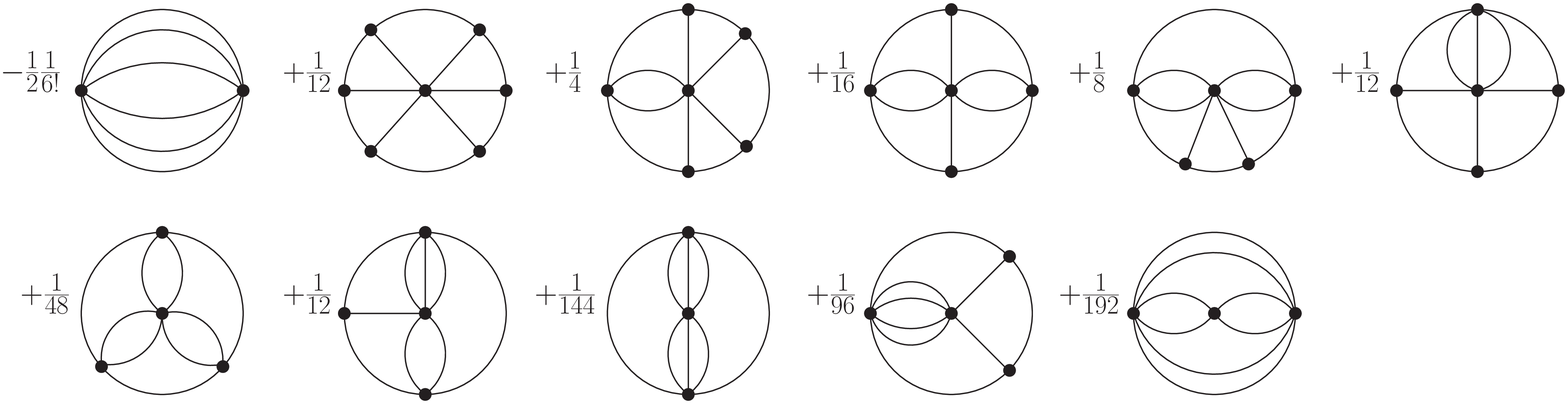}
\end{center}
\caption{\label{aa}6-loop diagrams with highest vertex $V_6$ and the basketball with $V_6$.}
\end{figure}
\par\begin{figure}[H]
\begin{center}
\includegraphics[width=14cm]{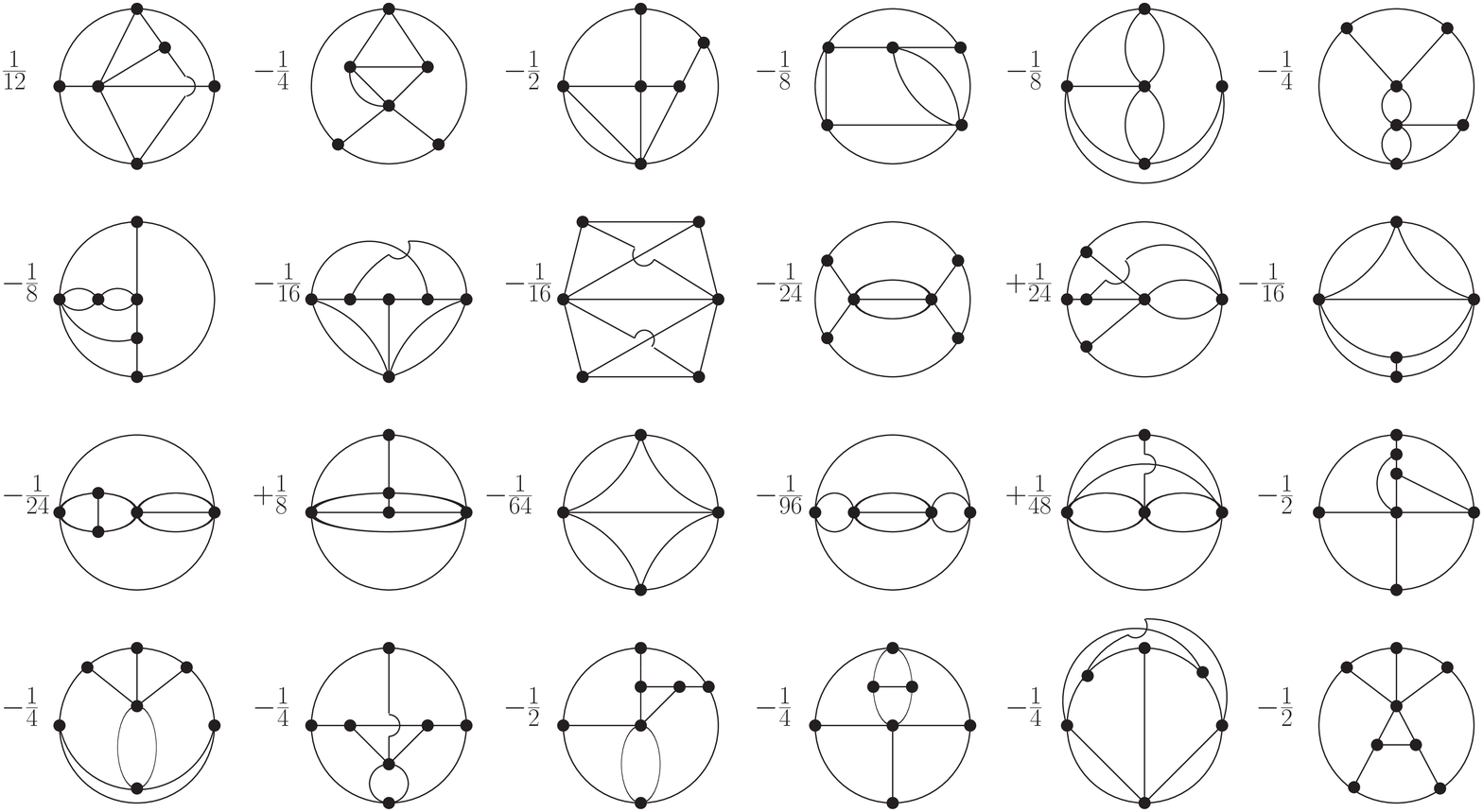}
\end{center}
\caption{\label{bb}6-loop diagrams with highest vertex $V_5$.}
\end{figure}
\par\begin{figure}[H]
\begin{center}
\includegraphics[width=14cm]{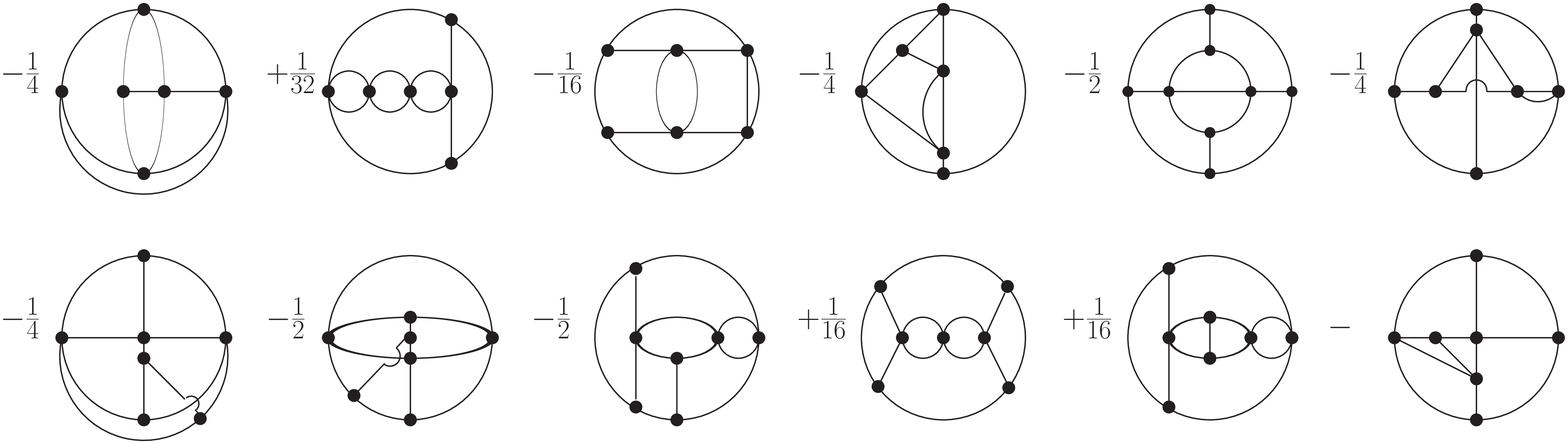}
\includegraphics[width=14cm]{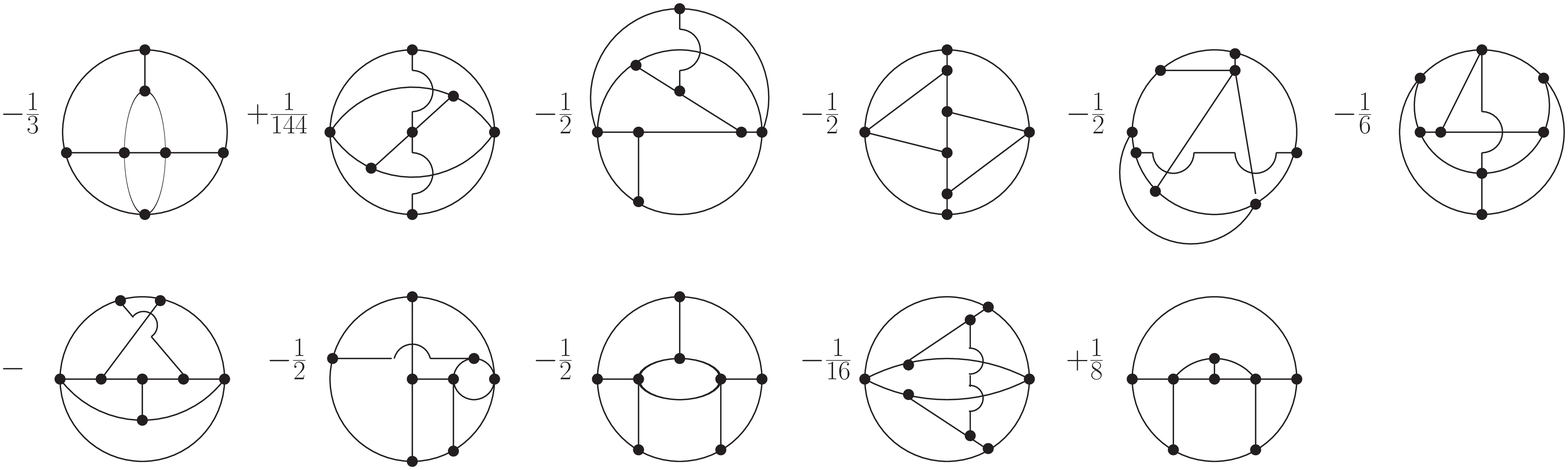}
\end{center}
\caption{\label{cc}6-loop diagrams with highest vertex $V_4$.}
\end{figure}
\par\begin{figure}[H]
\begin{center}
\includegraphics[width=12cm]{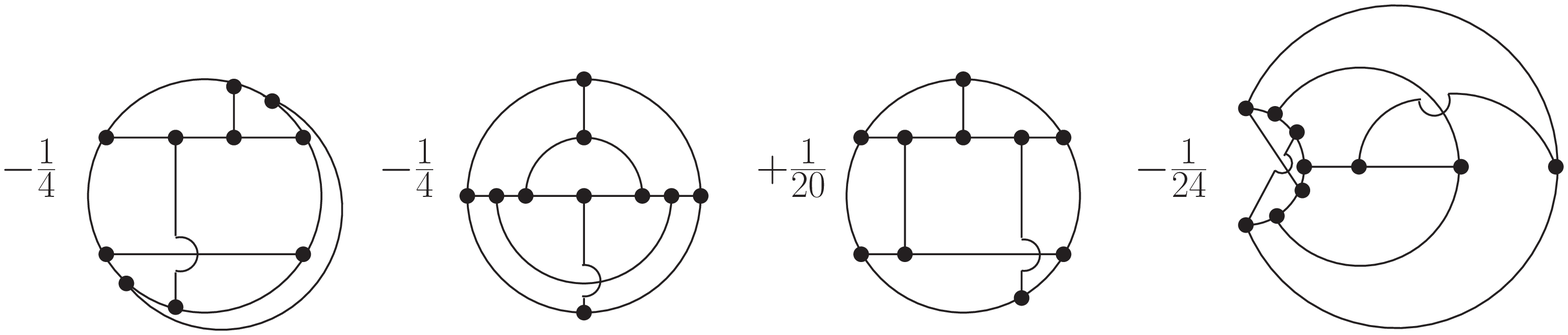}
\end{center}
\caption{\label{dd}6-loop diagrams with highest vertex $V_3$.}
\end{figure}

\section{Conclusions}

The $n$PI effective action at higher orders is a potentially useful tool to study nonequilibrium systems, like the quark gluon plasma and the early Universe.
In this paper, we have introduced a new method to calculate the $n$-Loop $n$PI effective action which does not require a Legendre transform and makes it possible to calculate the effective action at higher orders than was previously possible.
The key to our method is the introduction of a set of fictitious bare vertices which are used only as an organizational trick.
Using these fictitious vertices, we have shown that the $n$PI equations of motion and Schwinger-Dyson equations are equivalent to the order at which the truncated theory respects the symmetries of the original theory.
This result makes it possible to systematically construct the $n$-Loop $n$PI effective action directly from the sd equations, which are relatively easy to calculate.
The known results for the $n$-Loop $n$PI effective action with $n=4$ and $n=5$ can be obtained with comparatively little effort using our method, which provides a check of the procedure. In addition, we have used the technique to calculate the 6-Loop 6PI effective action, which is essentially impossible to obtain using the standard method employing Legendre transforms.

\appendix
\section{The 5-loop 5PI effective action - some details}
\label{5PIapp}

In this Appendix, we give some details of the calculation of the 5-loop 5PI effective action.

We show how to construct $I_4$ at the 2-Loop level.

\noindent \underline{explicit 2-loop}: We start with terms that are explicitly 2-loop in the sd equation, and set $V_j^0=V_j$ and $V_{j\ge 5}=0$. The survivors are the diagrams in parts (5,6,7,9,10,11,12,13,14) of Fig. \ref{SD4} with $V_j^0=V_j$.

\noindent \underline{explicit 1-loop}: Now, we look at terms that are explicitly 1-loop and set $V_{j\ge 5}=0$. These graphs are shown on the left side of Fig. \ref{fig18}. The numbers under the diagrams indicate the corresponding contribution in the sd equation (Fig. \ref{SD4}). The new graph which contains the fictitious vertex $V_5^0$ has coefficient $3/2-4=-1/2$ because there are contributions from the sd equation (Fig. \ref{extrasd}) and fcn$_4^\prime$ (Fig. \ref{extra}).
\par\begin{figure}[H]
\begin{center}
\includegraphics[width=17cm]{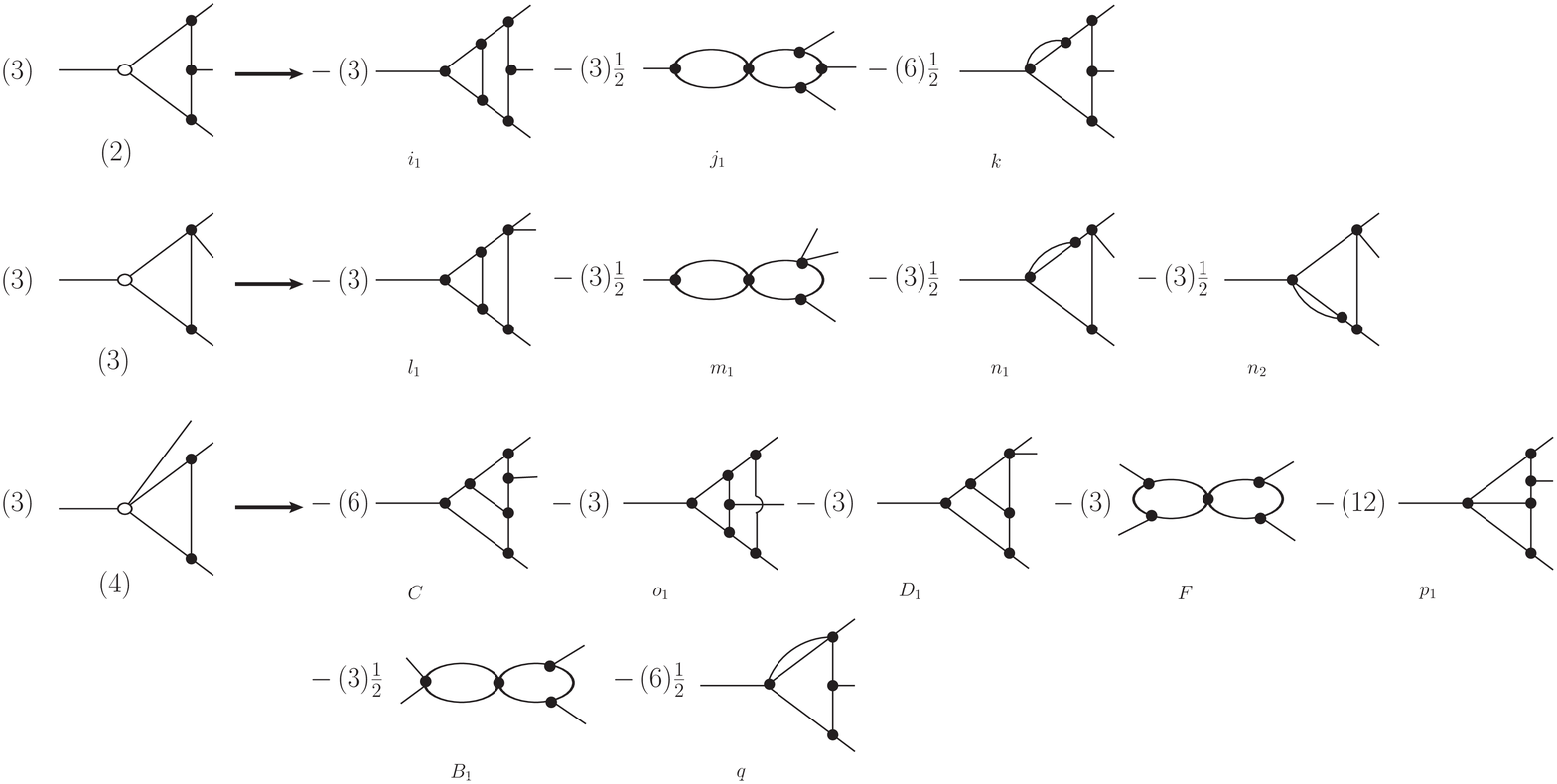}
\includegraphics[width=17cm]{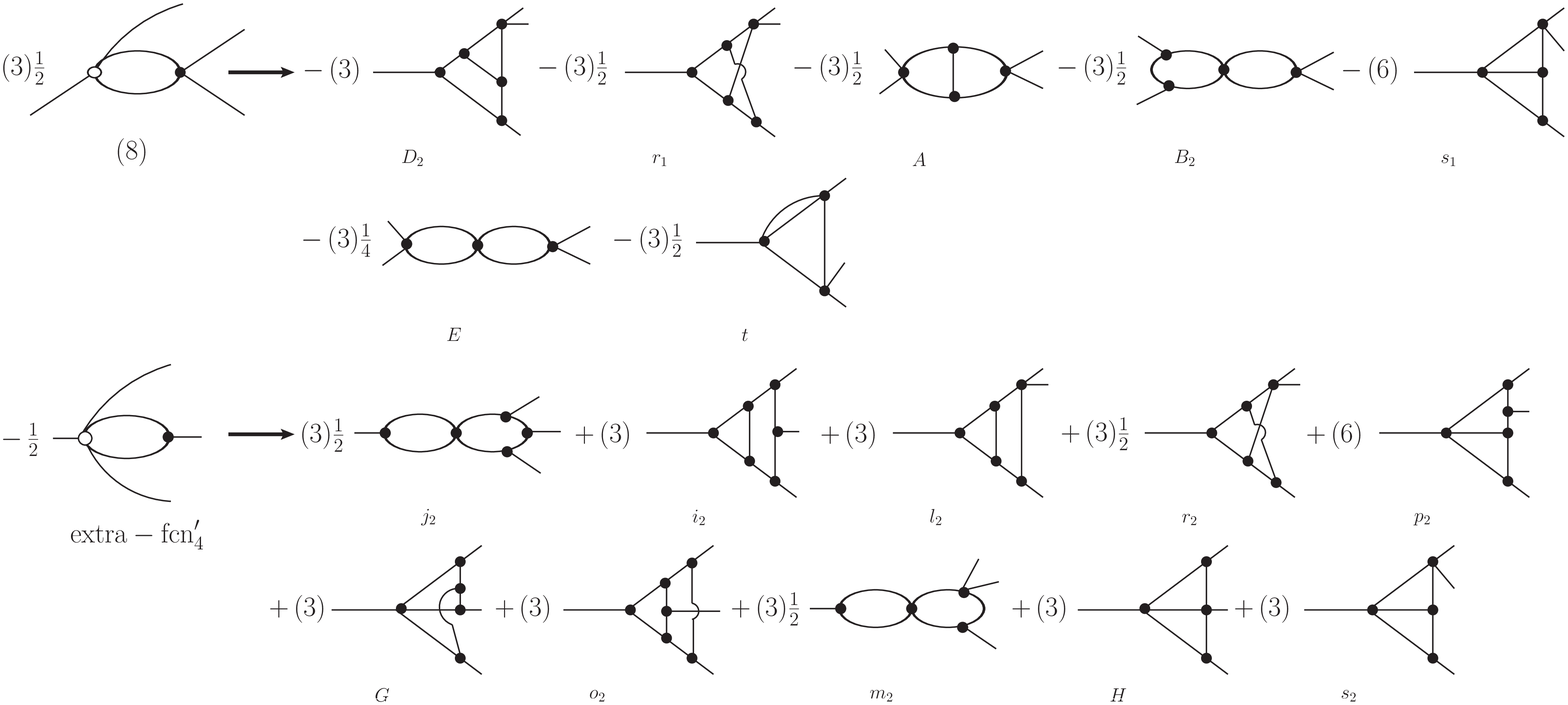}
\end{center}
\caption{\label{fig18} 2-loop terms from the explicit 1-loop part in $I_4$ with bare vertices removed using Fig. \ref{fig17}.}
\end{figure}

We remove bare vertices using (\ref{subbersd}) iterated to 1-Loop order with $V_{j\ge 5}=0$. The equations we obtain from (\ref{subbersd}) are shown in Fig. \ref{fig17}, and the 2-loop diagrams obtained by substituting these expressions into the explicit 1-loop diagrams in $I_4$ are shown on the right side of Fig. \ref{fig18}. The 1-loop diagrams can be ignored since they are, by construction, the same as in Fig. \ref{1loopv4}, and therefore produce the TARGET, EYEBALL, and LOOPY diagrams as in Sec. \ref{example44}.

\par\begin{figure}[H]
\begin{center}
\includegraphics[width=12cm]{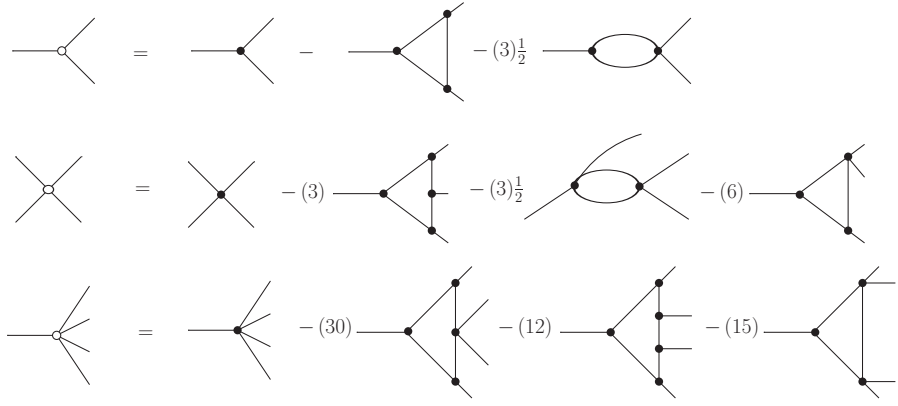}
\end{center}
\caption{\label{fig17} The result obtained from (\ref{subbersd}) which will be used to replace the bare vertices in the diagrams on the left side of Fig. \ref{fig18}.}
\end{figure}

We add the diagrams on the right side of Fig. \ref{fig18} and the explicit 2-loop terms [diagrams (5), (6), (7), (9), (10), (11), (12), (13), and (14) in Fig. \ref{SD4} with $V_j^0=V_j$].  The terms that cancel are
$i_1+i_2$, $j_1+j_2$, $(5)+k$, $l_1+l_2$, $m_1+m_2$, $(11)+(12)+n_1+n_2$, $o_1+o_2$, $(6)+p_1+p_2$, $(10)+q$, $r_1+r_2$, $(13)+s_1+s_2$, $(14)+t$.
The survivors are
A,
$B_1+B_2$, $C$, $D_1+D_2$, $E$, $F$, $(7)+G$, $(9)+H$ and are shown in Fig.  \ref{fig20}\footnote{In Fig. \ref{fig17}, the permutations of external legs are not symmetric. The numerical factor in brackets in front of each diagram indicates the number of permutations only.
However, for every set of graphs that cancels, each permutation of external legs cancels individually. For sets of graphs that do not cancel, every possible permutation of external legs is produced. Both of these results are guaranteed by the proof in Sec. \ref{sdeom}.}.
\par\begin{figure}[H]
\begin{center}
\includegraphics[width=12cm]{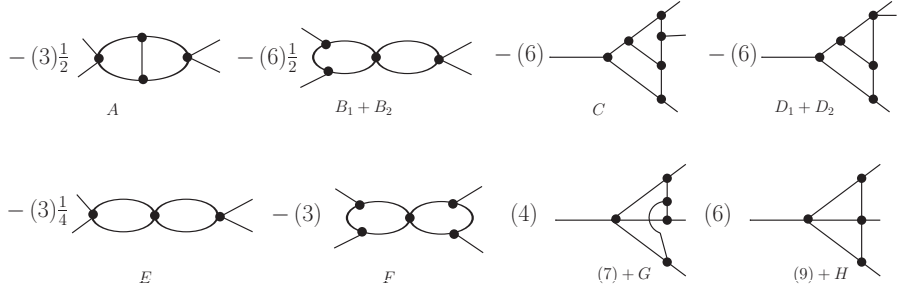}
\end{center}
\caption{\label{fig20} 2-loop contributions to $I_4$. The labels under the diagrams indicate the corresponding pieces of Figs. \ref{fig18} and \ref{SD4}.}
\end{figure}
The last step is to join the legs of each graph in Fig. \ref{fig20} and calculate the symmetry factor using (\ref{calS}). The first two diagrams in Fig. \ref{fig20} both give the same contribution to the effective action: the diagram 4A in Fig. \ref{fig:fig17}. For the first graph, the symmetry factor is $-3/2\cdot (1/3)\cdot (1/4!) = -1/48$, and for the second graph we have $-3\cdot (1/3)\cdot (1/4!) = -1/24$. Summing these factors, we obtain -1/16, which agrees with Fig. \ref{fig:fig17}. The last 6 diagrams give, in order, the graphs labeled 2B, 3D, 5A, 3A, BEAN, and PEA in Fig. \ref{fig:fig17}.

Next, we construct $I_3$ at the 3-Loop level.

$I_3$ is given by the terms (2,4,5,6,7,8) in Fig. \ref{SD3}, plus the additional terms in Fig. \ref{extrasd} which contain the bare vertex $V_5^0$, minus the terms in fcn$_3^\prime$ in Fig. \ref{extra}, with $V_4$=0.
We use (\ref{subbersd}) to iterate the sd equations to 2-Loop level, setting $V_{j\ge 4}=0$. This procedure produces the results in Fig. \ref{fig17} with $V_4=0$, plus the 2-loop diagrams in Fig. \ref{fig:50}. As mentioned in the discussion about Fig. \ref{fig18}, we do not need to separate graphs that correspond to different permutations of external legs, since we will join legs to obtain the corresponding contribution to the effective action. In Fig. \ref{fig:50}, we do not indicate contributions to the numerical factor from permutations of external legs.
\par\begin{figure}[H]
\begin{center}
\includegraphics[width=10cm]{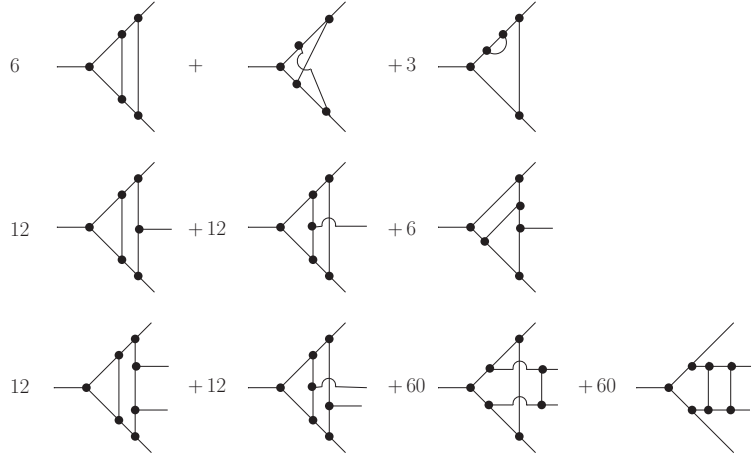}
\end{center}
\caption{\label{fig:50} 2-loop contributions to $V_3$, $V_4$, and $V_5$ obtained from (\ref{subbersd}) with $V_{j\ge 4}=0$. }
\end{figure}

\noindent \underline{explicit 3-loop:} Setting $V_j^0=V_j$ and $V_{j\ge 4}=0$, there are no surviving terms in $I_3$.

\noindent \underline{explicit 2-loop:} We take terms in $I_3$ that are explicitly 2-loop and set $V_{j\ge 4}=0$. Then, we replace bare vertices using the 1-Loop  expressions in Fig. \ref{fig17}.

\noindent \underline{explicit 1-loop:} We take terms in $I_3$ that are explicitly 1-loop and set $V_{j\ge 4}=0$. Then, we replace bare vertices using the 2-Loop  expressions in Figs. \ref{fig17} and \ref{fig:50}.

The 1- and 2-loop graphs that are produced by this procedure can be ignored, since they reproduce the MERCEDES and TWISTED diagrams obtained previously (Sec. \ref{example44}). After all cancellations have been identified, the surviving 3-loop diagrams are shown in Fig. \ref{fig21}. Joining the legs produces the graphs labeled 1A and 1B in Fig. \ref{fig:fig17}.
\par\begin{figure}[H]
\begin{center}
\includegraphics[width=7cm]{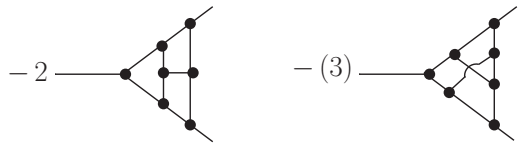}
\end{center}
\caption{\label{fig21} 3-loop diagrams in $I_3$.}
\end{figure}


\section{Formula for disconnected pieces of correlation functions}
\label{yunFormula}

We give a general expression for the function $\chi_j$ of the form
\bea
\label{side2}
\chi_j = \sum_k a^{[k]} \prod _{i=1}^{j-1} (V_i^c)^{f_i^{[k]}}\,.
\eea
The index $k$ represents different solutions to the equation
\bea
\label{options}
\sum_{i=1}^{j-1}i\,f^{[k]}_i =j\,, ~~j\ge 3\,.
\eea
The symmetry factor $a^{[k]}$ for each term is given by
\bea
\label{side3}
&& a^{[k]}=\prod_{i=1\big|f_i^{[k]}\ne 0}^{j-1} \alpha_i\,,~~~\alpha_i = \frac{1}{f_i^{[k]}!}\prod_{m=0}^{f_i^{[k]}-1}{\rm C}_{j^{\prime}_i-i m}^i\,,~~~
j^{\prime}_i=j-\sum_{l=0}^{i-1}l\,f_l^{[k]}\,,
\eea
where we have defined ${\rm C}_m^j\equiv m!/(j!(m-j)!)$.
We illustrate this formula with an example.
For $j=5$, the possible solutions to (\ref{options}) are
\bea
\label{n5solutions}
k=1:~~f^{[1]}_1=5,~f^{[1]}_2=0,~f^{[1]}_3=0,~f^{[1]}_4=0\,,\\
k=2:~~f^{[2]}_1=3,~f^{[2]}_2=1,~f^{[2]}_3=0,~f^{[2]}_4=0\,,\nonumber\\
k=3:~~f^{[3]}_1=1,~f^{[3]}_2=2,~f^{[3]}_3=0,~f^{[3]}_4=0\,,\nonumber\\
k=4:~~f^{[4]}_1=2,~f^{[4]}_2=0,~f^{[4]}_3=1,~f^{[4]}_4=0\,,\nonumber\\
k=5:~~f^{[5]}_1=1,~f^{[5]}_2=0,~f^{[5]}_3=0,~f^{[5]}_4=1\,, \nonumber\\
k=6:~~f^{[6]}_1=0,~f^{[6]}_2=1,~f^{[6]}_3=1,~f^{[6]}_4=0 \,. \nonumber
\eea

\begin{itemize}

\item For the $k=6$ solution in (\ref{n5solutions}), there are two nonzero values $f^{[6]}_2=1$ and $f^{[6]}_3=1$, which means $a^{[6]} = \alpha_2 \cdot \alpha_3$.

\item For the $\alpha_2$ term, we have $j_2^\prime = 5-f^{[6]}_1=5$ and $C^2_{5-2 m}= (5-2m)!/(2!(3-2m)!)$. The limits on the $m$ product are 0 to $f^{[6]}_2-1=0$, and therefore $m=0$ is the only term that contributes. The $m=0$ term is $C^2_{5}=10$. We multiply by a factor $1/f^{[6]}_2!$ = 1 and obtain $\alpha_2=10.$

\item For the $\alpha_3$ term, we have $j_3^\prime = 5-f^{[6]}_1-2f^{[6]}_2=3$ and $C^3_{3-3m}=(3-3m)!/(3!(-3m)!)$. The limits on the $m$ product are 0 to $f^{[6]}_3-1=0$, and therefore $m=0$ is again the only term that contributes, which gives $C^3_3=1$. We multiply by a factor $1/f^{[6]}_3!$ = 1 and obtain $\alpha_3=1.$

\item Multiplying these factors together, the result is $a^{[6]} =\alpha_2\cdot \alpha_3 =  10\cdot 1$.

\item Substituting into (\ref{side2}), the contribution to $\chi_5$ from the $k=6$ solution in (\ref{n5solutions}) is $10 D V^c_3$.

\end{itemize}

In the same way we calculate the $a^{[k]}$ for $k=1,2,3,4,5$. We give a summary of the results.

\bea
\begin{array}{lllll}
 k=5:~~& \alpha_1=5, & ~ \alpha_4=1,& ~ a^{[5]}&=5\,, \\
 k=4:~~& \alpha_1=10, & ~\alpha_3=1,& ~a^{[4]}&=10\,,\\
k=3:~~&   \alpha_1=5, & ~\alpha_2=3,& ~a^{[3]}&=15\,,\\
 k=2:~~&\alpha_1=10, & ~\alpha_2=1,& ~ a^{[2]}&=10\,,\\
 k=1:~~& \alpha_1=1, & ~ &~ a^{[1]}& =1\,.
 \end{array}
\eea
Combining these results, Eq. (\ref{side1}) for $j=5$ becomes
\bea
\label{n5res}
\langle \phi^5\rangle = V_5^c+(5) V_4^c \phi +(10) V_3^c \phi^2 +(10) V_3^c D +(10) D \phi^3 +(15) D^2 \phi +\phi^5\,.
\eea

In Sec. \ref{structure}, we divide the term $\chi_j$ into different pieces by defining
\bea
\chi_j = \chi_j^{(0)} + \chi_j^{(1)} + \chi_j^{(2)} + \chi_j^{(3)}\,.
\eea
We explain this notation below.
It is clear that (\ref{options}) always has three solutions which we write
\bea
&& k=1: ~~f^{[1]}_1=j\,, ~f^{[1]}_l=0,~l\ge 2 \,,\\
&& k=2: ~~f^{[2]}_1=j-2\,,~ f^{[2]}_2=1\,,~f^{[2]}_l=0,~l\ge 3 \,,\nonumber\\
&& k=3: ~~ f^{[3]}_1 = j-2d\,,~f^{[3]}_2=d\,,~f^{[3]}_l=0,~l\ge 3 ~{\rm and}~ j/2\ge d \ge 2\,.\nonumber
\eea
From (\ref{side3}), it is easy to see that $a^{[1]}=1$, $a^{[2]}=C_j^2$, and $a^{[3]} = j!/((j-2d)! d! 2^d)$. These three solutions give, respectively,
\bea
\label{chiDef}
&&\chi_j^{(0)} = \phi^j~~{\rm  for}~~ j\ge 3\,, \\
&&\chi_j^{(1)} = C_j^2 D \phi^{j-2}~~{\rm  for}~~ j\ge 3 \,,\nonumber\\
&& \chi_j^{(2)} = \frac{j!}{(j-2d)!d!2^d}D^d \phi^{j-2d}~~{\rm  for}~~j\ge 4 ~~{\rm and}~~j/2\ge d\ge 2\,.  \nonumber
\eea
The term $\chi_j^{(3)}$ is defined to be everything that is not contained in $\chi_j^{(0)}+ \chi_j^{(1)}+ \chi_j^{(2)}$.

\section{Bare vertex part of the effective action}
\label{bareSection}
In this section, we discuss how to calculate the part of the effective action that contains bare vertices.
We look at the example $V^{oo}_{j\ge 5}=0$. Using (\ref{scl}), (\ref{free}), (\ref{listDef}) and (\ref{chiDef}) we obtain
\bea
\label{getphi0}
&& \Gamma^{oo}_0 -\sum_{j=3}^4\frac{i}{j!} \chi_j^{(0)}V_j^{oo} = S_{cl}\,,\\
&& \Gamma^{oo}_1 -\sum_{j=3}^4\frac{i}{j!} \chi_j^{(1)}V_j^{oo} = \frac{i}{2} {\rm Tr} \,{\rm Ln}D^{-1}  +
\frac{i}{2} {\rm Tr}\,\big[ \left(D^0\right)^{-1} D\big]\,,\nonumber \\
&& -\sum_{j=3}^4\frac{i}{j!}\chi_j^{(2)}V_j^{oo} = -i{\rm EIGHT}\,, \nonumber\\
&& -\sum_{j=3}^4\frac{i}{j!}\chi_j^{(3)}V_j^{oo} = -\frac{i}{4!}(4)V_3^c\phi V_4^{oo}= -i{\rm EGG}_0^{a}\,,\nonumber\\
&& -\sum_{j=3}^4\frac{i}{j!}V_j^cV_j^{oo} = -\frac{i}{3!}V_3 D^3 V_3^{oo} -\frac{i}{4!} D^4[V_4 + 3(V_3 D V_3)]V_4^{oo} = -i{\rm EGG}_0^{b} -i {\rm BBALL}_0-i{\rm HAIR}\,.\nonumber
 \eea
The terms in the square bracket in the last line come from rewriting the connected vertex $V^c$ in terms of proper vertices. Adding the EGG contributions, we get the diagram with one effective bare vertex: ${\rm EGG}_0^{a}+{\rm EGG}_0^{b} = {\rm EGG}_0[V_3^0,V_3]$. Combining all contributions, we obtain the result for $\Gamma^0[\phi,D,V_j^{0},V_k]$ in Eq. (\ref{gamma0Def2}) with
\bea
\label{Phi0}
&& \Phi^0[V_j^0,V_k] = {\rm EIGHT} + {\rm EGG}_0 + {\rm HAIR} + {\rm BBALL}_0\,.
\eea
The diagrams denoted EIGHT, EGG$_0$, HAIR, and BBALL$_0$ are shown in Fig. \ref{fig:Phi0}.
Equation (\ref{Phi0}) is the usual result for the part of the effective action that contains bare vertices (see, for example, \cite{paper1}).

It is straightforward to calculate $\Gamma^0[\phi,D,V_j^{0},V_k]$ for a theory with fictitious vertices.
The classical action will contain additional terms [see Eq. (\ref{scl})].
The effective bare propagator and effective bare vertices are correspondingly modified [see Eq. (\ref{free})].
The 1-loop piece will have the same functional form; the only change is that it will depend on the modified effective bare propagator.
All of the graphs in Fig. \ref{fig:Phi0} will be present in the same form; the only change is that they now depend on the modified effective bare vertices. There will also be new contributions to $\Phi^0$. For the example $V^{oo}_5$ and $V^{oo}_6$ nonzero, the  new graphs are shown in Fig. \ref{fig:Phi0new}.

\section{$k_{\rm max}$ and basketballs}
\label{superBballSection}

In this Appendix, we show that in the $m$-Loop $n$PI effective action, the largest vertex that appears is $V_{k_{\rm max}}$ with $k_{\rm max}=m+1$, and the vertex $V_{m+1}$ appears only in the $m$-loop basketball diagram.

\ts

Using $I$ for the number of internal lines, $E$ for the number of external legs,  and $v_k$ for the number of $k$-point vertices, the standard topological relations are
\bea
\label{topo1}
m=I-\sum_{k=3}^n v_k +1\,,~~~~
 2I + E=\sum_{k=3}^n k\,v_k\,.
\eea
Eliminating $I$ and setting $E = 0$, we get
\bea
\label{topo}
m = 1+\sum_{k=3}^{k_{max}}\bigg(\frac{1}{2}k-1\bigg)v_k\,.
\eea
Our goal is to find $k_{\rm max}$ for fixed $m$. We note that every term in the sum in (\ref{topo}) is positive.

\ts

\noindent \underline{Case 1:} It appears that $k_{\rm max}$ corresponds to $v_k=0$ for $k\ne k_{\rm max}$ and $v_{k_{\rm max}}=1$. Substituting into (\ref{topo}), we obtain $k_{\rm max}=2m$. However, diagrams with only one vertex are type 1 tadpoles [see part ($b_1$) of Fig. \ref{superExLABEL}], and we know that tadpole graphs do not appear in the effective action (see Sec. \ref{structure}).

\ts

\noindent \underline{Case 2:} We consider the solution $v_k=0$ for $k\ne k_{\rm max}$ and $v_{k_{\rm max}}=2$, which corresponds to a $m$-loop basketball diagram. Substituting into (\ref{topo}), we obtain $k_{\rm max}=m+1$.

\ts

\noindent \underline{Case 3:} In order to conclude that $k_{\rm max}=m+1$ is the biggest solution for $k_{\rm max}$, we must check the case  $v_{k_{\rm max}}=1$ and $v_k\ne 0$ for some values $k<k_{\rm max}$.
We need to determine maximum number of legs from the vertices $V_{k\ne k_{\rm max}}$ that are available to connect with the lone $V_{k_{\rm max}}$ vertex without producing a tadpole graph. It is clear that no vertex can have two legs that connect to each other [to avoid creating a tadpole like the graph shown in part ($b_3$) of Fig. \ref{superExLABEL}], and each vertex must connect to at least two other vertices [to avoid creating a tadpole like the graph shown in part ($b_2$) of Fig. \ref{superExLABEL}]. Thus, the maximum number of legs from the vertices $V_{k\ne k_{\rm max}}$ that can connect to the $V_{k_{\rm max}}$ vertex is
\bea
\label{flowerMax2}
k_{\rm max}=\sum_{k=3}^{k_{\rm max}-1}(k v_{k} -2 v_{k})+2\,.
\eea
A graph that corresponds to Eq. (\ref{flowerMax2}) is given in Fig. \ref{proof}.
Rearranging (\ref{topo}) in the form
\bea
\label{flowerEqn}
m = 1+\frac{1}{2}\sum_{k=3}^{k_{\rm max}-1}\bigg(k-2\bigg)v_k+\frac{k_{\rm max}}{2}-1\,,
\eea
and substituting (\ref{flowerMax2}) into (\ref{flowerEqn}) we obtain $k_{\rm max}=m+1$, as in Case 2 above. This result appears to indicate that there is a large set of diagrams of the form shown in Fig. \ref{proof}, in addition to the $m$-loop basketball diagram, that contains the vertex $V_{k_{\rm max}=m+1}$. However, all diagrams of the form shown in Fig. \ref{proof} are flower topologies, which we know do not appear in the effective action (see Sec. \ref{structure}).
\par\begin{figure}[H]
\begin{center}
\includegraphics[width=9cm]{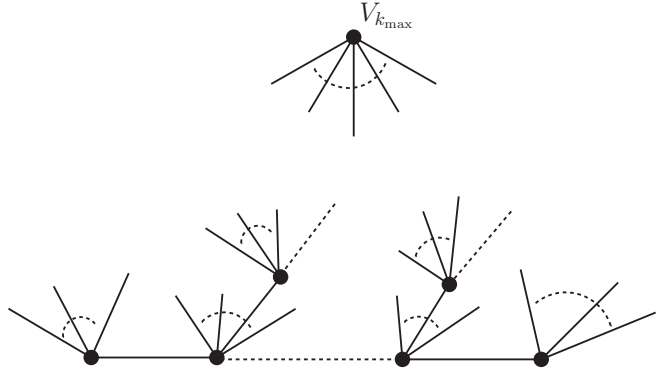}
\end{center}
\caption{\label{proof} The maximum number of legs that are available to be connected to $V_{k_{max}}$.}
\end{figure}

Since all terms in the sum in (\ref{topo}) are positive, it is clear that solutions that correspond to
$v_{k_{\rm max}}>2$,
or
$v_{k_{\rm max}}=2$ and $v_{k}\ne 0$ for some values $k<k_{\rm max}$,
will produce a smaller value of $k_{\rm max}$.
We conclude that in the $m$-Loop $n$PI effective action, the largest vertex that appears is $V_{k_{\rm max}}$ with $k_{\rm max}=m+1$, and the vertex $V_{m+1}$ appears only in the $m$-loop basketball diagram.

\section{Equivalence of the eom and sd equation for the self-energy}
\label{piSpecial}

The equation of motion for the 2-point vertex function obtained from the $n$-Loop $n$PI effective action can be rearranged to have the same form as the sd equation, without the use of fictitious vertices. In this Appendix, we prove this result.  The sd equation for the 2-point function is shown in Fig. \ref{sdPIeqnLABEL}.
\par\begin{figure}[H]
\begin{center}
\includegraphics[width=12cm]{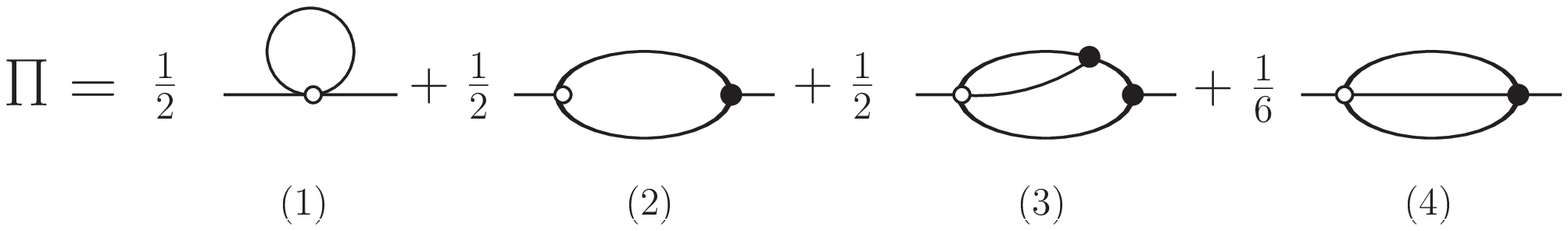}
\end{center}
\caption{\label{sdPIeqnLABEL}Schwinger-Dyson equation for the 2-point vertex. }
\end{figure}
\noindent The eom for the 2-point function is obtained from Eq. (\ref{2ptEx}). For the moment, we continue to use the abbreviated notation in which the indices which indicate the coordinates of each leg are suppressed.
Using this notation, the eom can be written
\bea
\label{appE1}
\Pi=\frac{1}{2}V_4^0 D + 2 \frac{1}{2!} V_3 D^2 V_3^0+ 2 \frac{1}{3!}V_4 D^3 V_4^0
- \sum_{j=3}^n\frac{1}{(j-1)!}V_j D^{j-1}V_j+\sum_i\Pi[{\rm diag}^{(i)}]\,.
\eea
The first three terms in this expression come from differentiating the EIGHT, EGG$_0$, and BBALL$_0$ diagrams in $\Phi^0$, respectively, (see Fig. \ref{fig:Phi0}). The first sum gives the contributions from the basketball diagrams in $\Phi^{\rm int}$ which we will call $\Pi$ basketballs. The second sum contains contributions from the HAIR diagram in $\Phi^0$ and all nonbasketball diagrams in $\Phi^{\rm int}$.
We can replace one of the vertices in each $\Pi$-basketball diagram using Eq. (\ref{EOM}). Rewriting (\ref{EOM}) as
\bea
\label{appE1b}
V_j=V_j^0+{\rm fcn}_j^\prime[V_l^0,V_k]+{\rm fcn}_j[V_k] =:V_j^0+ \sum_i{\rm f}_j[{\rm diag}^{(i)}]\,,
\eea
we obtain
\bea
\label{appE2}
&& \Pi=\frac{1}{2}V_4^0 D+ \frac{1}{2!} V_3 D^2 V_3^0 +\frac{1}{3!}V_4 D^3 V_4^0 + \sum_i{\rm EX}^{(i)}\,,\\
&& {\rm EX}^{(i)} = -\sum_{j=3}^n \frac{1}{(j-1)!}V_j D^{j-1}{\rm f}_j[{\rm diag}^{(i)}]  + \Pi[{\rm diag}^{(i)}]\nonumber\,.
\eea
The first three terms on the right side of the first line of (\ref{appE2}) are the diagrams labeled (1), (2), and (4) in Fig. \ref{sdPIeqnLABEL}. It is straightforward to show that EX$^{({\rm HAIR})}$  is the  diagram labeled (3) in Fig. \ref{sdPIeqnLABEL}. We show below that EX$^{(i)}$=0 for any diagram except for the HAIR diagram.

An arbitrary diagram with symmetry factor $S$, $I$ internal lines, and $v_k$ vertices $V_k$ for $k\ge 3$ can be written
\bea
\label{appE22}
{\rm diag}^{(i)} = S D^I \Pi_k V_k^{v_k}\,.
\eea
Using (\ref{EOMforce}), (\ref{2ptEx}), and (\ref{appE22}), we have
\bea
\label{appE3}
 && \Pi[{\rm diag}^{(i)}]= 2 I\,\bigg[S D^{I-1} \prod_k V_k^{v_k}\bigg] = 2 I \frac{1}{D}{\rm diag}^{(i)}\,,\\
&& {\rm f}_j[{\rm diag}^{(i)}] = j!D^{-j} \,\bigg[v_j S D^I V_j^{v_j-1}\prod_{k\ne j}V_k^{v_k}\bigg] = j! v_j D^{-j}\frac{1}{V_j} {\rm diag}^{(i)}\,.\nonumber
\eea
Substituting (\ref{appE3}) into the last line of (\ref{appE2}), we have
\bea
\label{appE4}
{\rm EX}^{(i)}=\left(-\sum_{j=3}^n j v_j + 2I\right){\rm diag}^{(i)}\frac{1}{D} = 0\,,
\eea
where we have used (\ref{topo1}) with $E=0$ in the last step.

The discussion above does not take into account the fact that Eq. (\ref{appE3}) can contain terms with  different topologies and permutations [see (\ref{proper}) and the discussion which follows this equation].
For an arbitrary diagram in the effective action, the contribution to the self-energy that is produced by opening one line can be different, depending on which line is opened.
We must show that each topology that is produced cancels individually. It is straightforward to see how this works.
Consider the example where diag$^{(i)}$ is taken to be the graph EIGHT4 in Fig. \ref{fig:fig17}. We consider the contribution to EX$^{({\rm EIGHT4})}$ from one line in the diagram and the two vertices this line attaches to, where the contribution to f$_j$ from these vertices is divided by the numerical factor $j$. If we can show that these contributions cancel, then it is clear that the contributions from any and all lines and their vertex partners cancel. Note that the vertex contribution must be divided by the factor $j$ because each vertex must partner with $j$ different lines.

We consider the case where the designated line is the horizontal line in the EIGHT4 diagram. This diagram is redrawn in Fig. \ref{fig:fig51}(a). The corresponding contribution to the self-energy is shown in part (b) of Fig. \ref{fig:fig51}. The contributions to the functions f$_5$ and f$_3$ from the vertices which attach to each end of the designated line are shown in parts (c) and (d), where the index $x$ indicates the leg of the vertex that was attached to the designated line in the original diagram in part (a). We substituting the graph in part (c) into the right side of the $\Pi$-basketball diagram that contains two $V_5$'s and the graph in part (d) into the right side of the $\Pi$-basketball diagram that contains two $V_3$'s. These two substitutions produce the two different permutations that are indicated by the factor (2) in front of the diagram in part (b). The numerical factors are $(2)\cdot 1/8-1/4!\cdot 1/8 \cdot 5!\cdot [1/5] -1/2! \cdot 1/8 \cdot 3!\cdot [1/3]=0$.
\par\begin{figure}[H]
\begin{center}
\includegraphics[width=15cm]{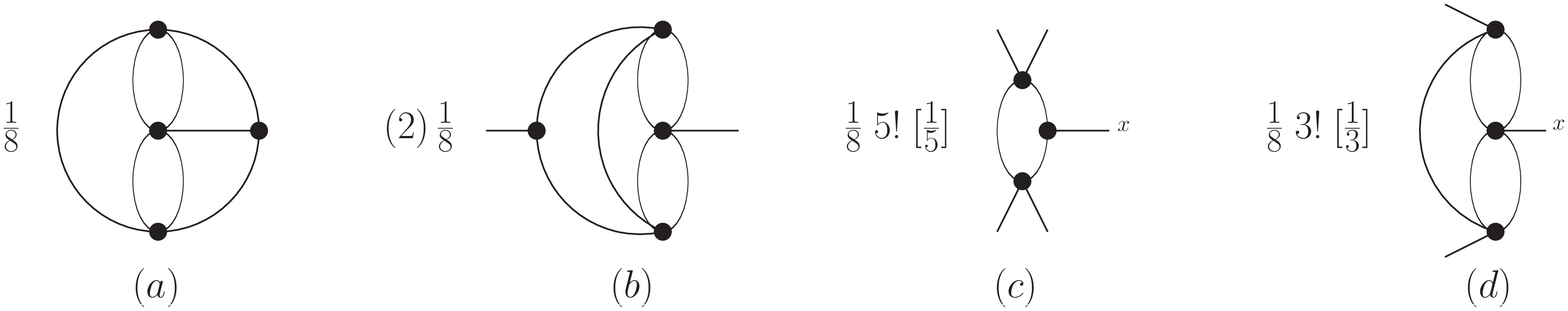}
\end{center}
\caption{\label{fig:fig51}The cancellation of one part of EX$^{({\rm EIGHT4})}$. The square brackets indicate the factors $1/j$ discussed in the text under Eq. (\ref{appE4}).}
\end{figure}
The procedure above can be applied to any line in any diagram.
It is easy to see that the numerical factors are always correct to produce a cancellation. In the first line of Eq. (\ref{appE3}), a factor $I$ is removed since only one line is differentiated.
In the second line of Eq. (\ref{appE3}), a factor $j\,v_j$ is removed because only one vertex is differentiated, and the contribution is divided by $1/j$ (so it can be used $j-1$ more times in partnership with the $j-1$ other lines that connect to it).
We obtain
\bea
\label{appE3l}
 && \Pi[{\rm diag}^{(i,l)}]=  2  \frac{1}{D}{\rm diag}^{(i,l)}\,,\\
 && {\rm f}_j[{\rm diag}^{(i,l)}] = (j-1)! D^{-j}\frac{1}{V_j} {\rm diag}^{(i,l)}\,,\nonumber
\eea
where the notation ${\rm diag}^{(i,l)}$ indicates that an arbitrary line labeled $(l)$ in the diagram labeled $(i)$ is considered.
Using these results, (\ref{appE4}) becomes
\bea
\label{appE4l}
{\rm EX}^{(i,l)}=(-1-1+2)\,{\rm diag}^{(i,l)}\frac{1}{D} = 0\,.
\eea
The two terms -1 in the equation above correspond to the two nonzero terms in the sum in (\ref{appE4}) which come from the two vertices that attach to the designated line.

Thus, we have proved that ${\rm EX}^{(i)}=0$ in (\ref{appE2}). Equivalently, we have shown that when we substitute (\ref{appE1b}) into the vertex on the right side of each $\Pi$ basketball in (\ref{appE1}), the sum of all terms produced by the functionals ${\rm f}_j$ cancel with the second sum in (\ref{appE1}).

We must also consider the term produced by the bare vertex from the first term on the right side of (\ref{appE1b}). This term produces the two $\Pi$-basketball topologies which have a bare $V_3^0$ and $V_4^0$ on the right side. However, the second and third terms on the right side of (\ref{appE1}) contain two graphs each, which are $\Pi$-basketball topologies with the bare vertex on the left and right sides. The result is that the graphs with bare vertices on the right side cancel, and we are left with the graphs labeled (2) and (4) in Fig. \ref{sdPIeqnLABEL}.

Note that the HAIR diagram contains a bare vertex $V_4^0$, and therefore the lines that attach to the bare vertex do not have partner contributions from a term of the form ${\rm f}_4^{\;({\rm HAIR},l)}$, which means that EX$^{({\rm HAIR})}\ne 0$. As mentioned above, it is straightforward to show that EX$^{({\rm HAIR})}$ is diagram (3) in Fig. \ref{sdPIeqnLABEL}.


\begin{thebibliography}{99}
\bibitem{deDom1}C. DeDominicis and P.C. Martin, J. Math. Phys. (N.Y.) {\bf 5}, 14 (1964).
\bibitem{deDom2}C. DeDominicis and P.C. Martin, J. Math. Phys. (N.Y.) {\bf 5}, 31 (1964).
\bibitem{norton} R.E. Norton and J.M. Cornwall, Ann. Phys. (N.Y.) {\bf 91}, 106  (1975).
\bibitem{cox} J. Berges and  J. Cox, Phys. Lett. {\bf B517}, 369 (2001)  - {\it arXiv:hep-ph/0006160}.
\bibitem{aartsNonEq0} G. Aarts and J. Berges, Phys. Rev. {\bf D64}, 105010, (2001) - {\it arXiv:hep-ph/0103049}.
\bibitem{smitNonEq} Alejandro Arrizabalaga, Jan Smit and Anders Tranberg, Phys. Rev. {\bf D72}, 025014  (2005)  - {\it Xiv:hep-ph/0503287}.
\bibitem{bergesNonEq}J. Berges, Sz. Borsanyi and C. Wetterich, Nucl. Phys. {\bf B727}, 244  (2005) - {\it arXiv:hep-ph/0505182}.
\bibitem{aartsNonEq1} Gert Aarts and Anders Tranberg,  Phys. Rev. {\bf D74}, 025004 (2006) - {\it  arXiv:hep-th/0604156}.
\bibitem{aartsNonEq2} Gert Aarts, Nathan Laurie and Anders Tranberg, Phys. Rev. {\bf D78}, 125028 (2008) - {\it  arXiv:0809.3390}.
\bibitem{edQED} M.E. Carrington and E. Kovalchuk, Phys. Rev. {\bf D77}, 025015 (2008) - {\it arXiv:0709.0706}.
\bibitem{edQCD} M.E. Carrington and E. Kovalchuk, Phys. Rev. {\bf D80}, 085013 (2009) - {\it arXiv:0906.1140}.
\bibitem{EK4} M.E. Carrington and E. Kovalchuk, Phys. Rev. {\bf D81}, 065017 (2010) - {\it arXiv:0912.3149}.
%
\bibitem{bergesConvg}J. Berges, Sz. Borsanyi, U. Reinosa and J. Serreau, Phys. Rev. {\bf D71},  105004 (2005) - {\it arXiv:hep-ph/0409123}.
\bibitem{bergesReview} J. Berges, in {\it Introduction to Nonequilibrium Quantum Field Theory}, edited by M. Bracco, M. Chiapparini, E. Ferreira and T. Kodama, AIP Conf. Proc. {\bf 739}, 3 (AIP, New York, 2005) - {\it arXiv:hep-ph/0409233}.
\bibitem{bergesSEWM2004} J. Berges and J. Serreau, in {\it Proceedings of Strong Electroweak Matter}, edited by  K.J. Eskola, (World Scientific, Singapore, 2004) - {\it arXiv:hep-ph/0410330}.
%
%
\bibitem{calzettaReview} E. Calzetta, Int. J. Theor. Phys. {\bf 43}, 767  (2004) -  {\it arXiv:hep-ph/0402196}.
\bibitem{julienReview} U. Reinosa and J. Serreau, Ann. Phys. (N.Y.) {\bf 325}, 969 (2010) - {\it arXiv:0906.2881}
%
\bibitem{baier2PI} G. Aarts, D. Ahrensmeier, R. Baier, J. Berges and J. Serreau, Phys. Rev. {\bf D66}, 045008 (2002) - {\it hep-ph/0201308}.
\bibitem{vanHees3} H. van Hees and J. Knoll Phys. Rev. {\bf D66}, 025028 (2002)  - {\it arXiv:hep-ph/0203008}.
%
\bibitem{edQED-2pi} M.E. Carrington and E. Kovalchuk, Phys. Rev. {\bf D76}, 045019  (2007) - {\it arXiv:0705.0162}.
\bibitem{julienSym} U. Reinosa and J. Serreau, JHEP, {\bf 0711}, 097 (2007) - {\it arXiv:0708.0971}.
\bibitem{calzettaSym} J. Peralta-Ramos and E. Calzetta, J. Phys.: Condens. Matter {\bf 21}, 215601 (2009) - {\it	 arXiv:0811.2765v2}.
%
\bibitem{vanHees1} H. van Hees and  J. Knoll, Phys. Rev. {\bf D65}, 025010 (2001) - {\it arXiv:hep-ph/0107200}
\bibitem{vanHees2} H. van Hees and J. Knoll, Phys. Rev. {\bf D65}, 105005 (2002)  - {\it  arXiv:hep-ph/0111193}.
%
\bibitem{reinosaRenorm1} J. Berges, S. Borsanyi, U. Reinosa and J. Serreau, Ann. Phys. (N.Y.) {\bf 320}, 344 (2005) - {\it arXiv:hep-ph/0503240}.
\bibitem{reinosaRenorm2} U. Reinosa and J. Serreau, JHEP {\bf 0607}, 028 (2006) - {\it arXiv:hep-th/0605023}.
\bibitem{smit} A. Arrizabalaga and J. Smit, Phys. Rev. {\bf D66}, 065014 (2002) - {\it arXiv:hep-ph/0301093}.
\bibitem{HZ} M.E. Carrington, G. Kunstatter and H. Zaraket, Eur. Phys. J. {\bf C42}, 253 (2005) - {\it arXiv:hep-ph/0309084}.

%
\bibitem{berges} J. Berges, Phys. Rev. {\bf D70}, 105010 (2004) - {\it  arXiv:hep-ph/0401172}.
\bibitem{MEC} M.E. Carrington, Eur. Phys. J. {\bf C35}, 383 (2004) - {\it arXiv:hep-ph/0401123}.
\bibitem{paper1} M.E. Carrington and Yun Guo, Phys. Rev. {\bf D83}, 016006 (2011) - {\it arXiv:1010.2978}.
\bibitem{jaxo} D. Binosi and L. Theussl, Comput. Phys. Commun. {\bf 161}, 76 (2004) - {\it arXiv:hep-ph/0309015}.
\bibitem{cvitanovic} P. Cvitanovi\'{c}, B. Lautrup and R.B. Pearson, Phys. Rev. {\bf D18}, 1939 (1978).
\bibitem{kajantie} K. Kajantie, M. Laine and Y. Schr\"{o}der, Phys. Rev. {\bf D65}, 045008 (2002) - {\it arXiv:hep-ph/0109100}.
\bibitem{alkoferALG}R. Alkofer, M. Q. Huber and K. Schwenzer, Comput. Phys. Commun. {\bf 180}, 965 (2009).



\end{thebibliography}
\end{document}